\documentclass[manuscript]{acmart}

\AtBeginDocument{%
  \providecommand\BibTeX{{%
    \normalfont B\kern-0.5em{\scshape i\kern-0.25em b}\kern-0.8em\TeX}}}

\usepackage{graphicx}
\usepackage{mathtools}
\usepackage{booktabs}
\usepackage{multirow}
\usepackage{enumitem}
\usepackage{array}
\usepackage{hyperref}

\begin{document}

\title[Chatting with a Learning Analytics Dashboard]{Chatting with a Learning Analytics Dashboard: The Role of Generative AI Literacy on Learner Interaction with Conventional and Scaffolding Chatbots}

\author{Yueqiao Jin}
\affiliation{%
  \institution{Monash University}
  \country{Australia}
}

\author{Kaixun Yang}
\affiliation{%
  \institution{Monash University}
  \country{Australia}
}

\author{Lixiang Yan}
\affiliation{%
  \institution{Monash University}
  \country{Australia}
}

\author{Vanessa Echeverria}
\affiliation{%
  \institution{Monash University}
  \country{Australia}
}

\author{Linxuan Zhao}
\affiliation{%
  \institution{Monash University}
  \country{Australia}
}

\author{Riordan Alfredo}
\affiliation{%
  \institution{Monash University}
  \country{Australia}
}

\author{Mikaela Milesi}
\affiliation{%
 \institution{Monash University}
 \country{Australia}}

\author{Jie Fan}
\affiliation{%
 \institution{Monash University}
 \country{Australia}}

\author{Xinyu Li}
\affiliation{%
 \institution{Monash University}
 \country{Australia}}
\email{xinyu.li1@monash.edu}

\author{Dragan Gašević}
\affiliation{%
  \institution{Monash University}
  \country{Australia}
}

\author{Roberto Martinez-Maldonado}
\affiliation{%
  \institution{Monash University}
  \country{Australia}
}

\renewcommand{\shortauthors}{Jin et al.}

\begin{abstract}

Learning analytics dashboards (LADs) simplify complex learner data into accessible visualisations, providing actionable insights for educators and students. However, their educational effectiveness has not always matched the sophistication of the technology behind them. Explanatory and interactive LADs, enhanced by generative AI (GenAI) chatbots, hold promise by enabling dynamic, dialogue-based interactions with data visualisations and offering personalised feedback through text. Yet, the effectiveness of these tools may be limited by learners’ varying levels of GenAI literacy, a factor that remains underexplored in current research. This study investigates the role of GenAI literacy in learner interactions with conventional (reactive) versus scaffolding (proactive) chatbot-assisted LADs. Through a comparative analysis of 81 participants, we examine how GenAI literacy is associated with learners’ ability to interpret complex visualisations and their cognitive processes during interactions with chatbot-assisted LADs. Results show that while both chatbots significantly improved learner comprehension, those with higher GenAI literacy benefited the most, particularly with conventional chatbots, demonstrating diverse prompting strategies. Findings highlight the importance of considering learners’ GenAI literacy when integrating GenAI chatbots in LADs and educational technologies. Incorporating scaffolding techniques within GenAI chatbots can be an effective strategy, offering a more guided experience that reduces reliance on learners’ GenAI literacy.
  
\end{abstract}





\maketitle

\vspace{-10pt}
\section{Introduction}
As educational technology rapidly evolves, learning analytics dashboards (LADs) are gaining increasing attention for their potential to transform education by leveraging data to enhance learning and teaching practice \cite{verbert2020learning,sahin2021visualizations}. These dashboards generally aim to simplify complex learner data into more accessible visualisations, offering educators, students, and other stakeholders insights that can potentially be meaningful and actionable \cite{kim2016effects,Pokhrel_2021}. By presenting information in a more visually digestible format, LADs aim to help learners and educators track progress \cite{Kaliisa_2024}, identify areas of strength and weakness \cite{Yousef_2021}, support informed decision-making \cite{Susnjak_2022}, and assist with timely interventions \cite{pinargote2024automating}.

Yet, despite significant advancements in learning analytics, and the development of LADs using state-of-the-art technologies, such as artificial intelligence (AI) and multimodal sensory data \cite{ochoa_multimodal_2022}, their educational \textit{effectiveness} has not always been matched with the sophistication of these technologies \cite{Kaliisa_2024}. Indeed, recent concerns have been raised about whether the increasing complexity of educational data, driven by advancements in data collection and processing technologies, may overwhelm learners and educators \cite{echeverria2018driving, corrin2018evaluating}. This complexity necessitates a growing need to ensure that visualisations remain comprehensible and do not contribute to cognitive overload \cite{Ramaswami_2022}.


A potential pathway to enhancing learners' comprehension of complex visualisations is the use of explanatory and interactive analytics that could serve to scaffold students' interactions with LADs \cite{echeverria2018driving, yan2024genai}. The rapid advancements in generative artificial intelligence (GenAI) tools, now equipped with multimodal capabilities, have made it increasingly feasible to utilise GenAI-powered conversational agents, commonly referred to as GenAI chatbots, to enrich LADs with interactive features and to transit from exploratory (focused on identifying patterns and generating hypotheses) to explanatory (aimed at explaining causes and providing deeper insights) analytics \cite{ooi2023potential,echeverria2018exploratory,sahin2021visualizations}. Studies have demonstrated that these chatbots can support learners' comprehension and help reduce their cognitive load by providing on-demand, real-time, and personalised explanations and feedback \cite{okonkwo2021chatbots, Kuhail_2022}. In addition, by integrating retrieval-augmented generation (RAG) methodologies, LADs can evolve from a one-way information delivery system to a conversational platform \cite{verbert2020learning,yan2024genai}. This transformation may allow learners to communicate and collaborate with the dashboard, access underlying resources and receive tailored guidance. Such interaction holds the potential to foster a more engaging and supportive learning environment, enabling learners to ask questions, seek clarifications, and receive immediate, contextually relevant responses \cite{verbert2020learning}. Additionally, by incorporating an agentic system design in LADs \cite{yan2024vizchat}, \textit{conventional chatbots} (that passively answer learners' questions) can also be transformed into \textit{scaffolding chatbots} (that actively guide learners through the key components of LADs), potentially enhancing their understanding and facilitating learning \cite{gibbons2002scaffolding, kim2018effectiveness}.

While the integration of various types of GenAI chatbots into LADs holds great promise, the extent to which students can engage fairly and effectively with these tools largely depends on their level of GenAI literacy \cite{Alzubi_2024}. GenAI literacy refers to the ability to understand, interact with, and effectively use GenAI-driven tools and technologies \cite{Annapureddy_2024}. As these tools become more prevalent in educational settings, it is crucial to consider how varying levels of GenAI literacy among learners might influence their ability to benefit from GenAI-powered conversational agents \cite{Kuhail_2022}. For instance, learners with high GenAI literacy are likely to navigate and benefit from chatbot-assisted LADs more efficiently, making the most of the personalised feedback and real-time interactions these tools offer \cite{Annapureddy_2024, Bozkurt_2024}. Conversely, learners with lower levels of GenAI literacy may struggle to fully engage with \textit{conventional chatbots} (those which reactively respond to questions, such as regular ChatGPT), potentially leading to frustration or underutilisation of the available resources if they are unsure how to formulate questions or unaware of the types of queries the agent can effectively respond to \cite{Lyu_2024, Alzubi_2024}. In contrast, a \textit{scaffolding chatbot} (which actively prompt learners with guiding questions) might lower the dependence on their GenAI literacy, considering the evident effectiveness of scaffolding techniques \cite{kim2018effectiveness}. Consequently, it is essential to consider learners' GenAI literacy when integrating GenAI-powered learning interventions.

Despite the emergence of many position pieces that emphasise the benefits and challenges of GenAI in learning analytics and LADs \cite{yan2024genai,cukurova2024interplay,hennessy2024bjet,khosravi2023generative}, there is a notable gap in empirical research examining the actual effectiveness of different types of chatbot-assisted LADs (e.g., conventional versus scaffolding chatbots) in enhancing learners' comprehension of learning insights. Likewise, the impact of GenAI literacy on learners' comprehension when using chatbot-assisted LADs also remains underexplored. Additionally, there is limited knowledge about the cognitive processing patterns exhibited by learners with varying GenAI literacy levels when interacting with these chatbots.

To address these gaps, this paper presents a study involving 81 learners, comparing the effectiveness of two types of chatbot-assisted LADs—\textit{conventional/reactive} and \textit{scaffolding/proactive}—to enhance learners' comprehension of complex visualisations. We then investigated the association between learners' GenAI literacy and their comprehension when interacting with these GenAI chatbots. It also examines how learners' GenAI literacy relates to their comprehension and cognitive processing patterns during interactions with these chatbots. The research aims to provide empirical insights into the role of GenAI literacy in effectively using chatbot-assisted LADs, informing the design of more inclusive educational technologies for all learners.

\section{Background}
\subsection{Learning Analytics Dashboard and Visualisations}

LADs are instrumental in aiding educational stakeholders in the ongoing assessment of teaching and learning methodologies \cite{sahin2021visualizations}. Comprising visual depictions and analyses of educational data, LADs encourage users to identify trends and address queries within the data \cite{verbert2020learning}. Nevertheless, LADs face significant constraints, including the presentation of intricate visualisations, ineffective insight communication \cite{corrin2018evaluating}, and an incongruity with teachers' pedagogical requirements \cite{Kaliisa2022}. To mitigate these challenges, the focus of research has pivoted to the development of \textit{explanatory LADs}. These are tailored to visually guide users to crucial insights contextualised with information from the learning design, thereby aligning more closely with the pedagogical needs of stakeholders \cite{echeverria2018driving}. Embracing information visualisation techniques and data storytelling principles, these explanatory LADs utilise clear titles to communicate main messages, narratives with text explanations, and visual markers such as arrows and colour-coded information to highlight specific data points \cite{echeverria2018driving,fernandez2022beyond}. Studies have demonstrated that these dashboards effectively direct attention to significant pedagogical insights \cite{echeverria2018driving,martinez2020data}, promote profound reflective practice, and assist both teachers and students in comprehending their learning processes \cite{echeverria2018driving,fernandez2022beyond}. Furthermore, they are particularly beneficial for teachers with limited visual literacy, facilitating an easier understanding of the information presented \cite{pozdniakov2023teachers}. Such augmentations also effectively enhance learners' comprehension of the key insights compared to conventional visualisations \cite{shao2024data}. Despite these advantages, a notable challenge remains: the automation of translating contextual information from learning activities into visual elements. This task frequently involves a co-design process where designers, researchers, and educators collaboratively define pedagogical intentions and align them with visual elements \cite{echeverria2018exploratory,martinez2020data,fernandez2022beyond,pozdniakov2023teachers}. Recent advancements have attempted to simplify this process by having educators specify pedagogical rules, although the actual translation into visual elements continues to rely on designers and researchers \cite{fernandez2024data}. In parallel, other solutions are beginning to utilise unsupervised learning and natural language processing (NLP) to automatically generate student narratives, thus enabling academic advisors to better understand student success and risk behaviours \cite{Ahmad2020, li2024we}.

\subsection{GenAI in Education}

Recent strides in GenAI have markedly expanded the horizons for creating educational content and interacting with educational data \cite{bahroun2023transforming, yan2024promises}. Large Language Models (LLMs) such as GPT and LLaMA have shown potential in autonomously generating instructional materials and practice quizzes, ushering in a new era in educational content creation \cite{yan2024practical, cukurova2024interplay, khosravi2023generative}. Concurrently, innovations like OpenAI's Whisper have improved the accessibility of learning materials by enabling automatic transcription of lecture recordings, catering to a variety of preferences and needs \cite{gris2023evaluating}. Additionally, image-generating diffusion models such as Midjourney and DALL-E have enriched multimedia learning and fostered creativity in art-focused STEAM education by translating text into compelling visuals \cite{lee2023prompt}. The advent of multimodal models like GPT-4o, which adeptly interpret both textual and visual information, represents a significant leap forward in providing contextually relevant explanations \cite{achiam2023gpt}. These advancements point to a future where the communication of insights to learners and educators is increasingly interactive, varied, and effective \cite{yan2024genai}.

Incorporating multimodal GenAI into LADs holds promise for improving explanatory capabilities through interactive dialogues \cite{yan2024vizchat, fernandez2024data}. Although chatbots have been used in various educational settings, such as intelligent tutoring systems and feedback mechanisms \cite{okonkwo2021chatbots}, their integration within LADs has faced challenges due to the difficulty of achieving high-quality, zero-shot understanding of visual language \cite{ren2023visual}. Effective explanations in LADs necessitate contextual grounding within specific learning designs to support decision-making and encourage deep reflection \cite{verbert2020learning, sahin2021visualizations}, which calls for advanced multimodal AI models capable of comprehending both natural and visual language. Moreover, the provision of accurate and contextually relevant explanations by GenAI in LADs relies on robust supporting infrastructures. Despite their potential, GenAI models often produce well-articulated but inaccurate outputs, a phenomenon known as "hallucination" \cite{ji2023survey}. To mitigate this, RAG techniques, which focus content generation on contextually pertinent materials \cite{shuster2021retrieval}, have gained traction. This involves encoding relevant information into vector embeddings that encapsulate semantic or relational meanings \cite{mikolov2013distributed}, allowing for dynamic retrieval during AI interactions through semantic searches using metrics like cosine similarity, thereby reducing hallucinations and enhancing domain-specific accuracy \cite{siriwardhana2023improving}. Furthermore, GenAI models can produce irrelevant responses if user prompts lack sufficient detail \cite{white2023prompt}. Since expecting all learners and educators to master the skill of effective prompt crafting is impractical, the deployment of AI agents, autonomous entities that refine prompts to achieve specific goals, emerges as a promising solution \cite{wu2023autogen}. By integrating multimodal GenAI with RAG and autonomous AI agents, it becomes feasible to develop context-aware chatbots for LADs that deliver precise, contextually grounded explanations of visualisations, thereby markedly enriching the educational experience for both learners and educators.

LADs integrated with agentic chatbots embedded with scaffolding techniques can be promising to deliver a more structured and engaging learning experience \citep{wu2023autogen, park2023generative}. Scaffolding, an educational strategy with a strong theoretical grounding, entails decomposing complex information into manageable chunks, posing guiding questions, and providing feedback to enhance learner comprehension of a specific subject \cite{gibbons2002scaffolding}. This can promote self-regulated learning by fostering metacognitive processes, encouraging monitoring behaviours, potentially improving learning outcomes \cite{wen2024learning, lim2023effects}. Empirical research underscores its effectiveness in aiding learners to achieve a deeper understanding and mastery by offering contextual guidance during the knowledge and skill development process \cite{gibbons2002scaffolding, kim2018effectiveness}. Therefore, the integration of scaffolding chatbots into LADs could potentially enhance learners' comprehension more efficaciously than conventional chatbots that simply respond to queries. While this method is promising and innovative system frameworks are currently under development \cite{yan2024vizchat, ma2023demonstration}, empirical evidence on the efficacy of these chatbots in improving learners' understanding of key insights presented within LADs remains limited. This gap informs our first research question: \textbf{RQ1}: To what extent do conventional and scaffolding GenAI chatbots enhance learners' \textit{comprehension} of key insights presented by visualisations in LADs?

\subsection{GenAI Literacy}

GenAI literacy is crucial to understanding how students interact with and benefit from different types of chatbots in educational settings. While AI literacy broadly encapsulates competencies such as critically evaluating AI technologies, communicating and collaborating effectively with AI, and using AI tools ethically in various contexts \cite{Long_2020, Ng_2021}, GenAI literacy addresses the specific skills required for engaging with generative models like ChatGPT \cite{Zhao_2024}. It is important to distinguish between general AI literacy and GenAI literacy, as existing frameworks often fail to cover the specific competencies needed to effectively utilise GenAI tools \cite{Annapureddy_2024}. For instance, \citet{Zhao_2024} suggests that GenAI literacy should include pragmatic, safety, reflective, socio-ethical, and contextual understandings. \citet{Bozkurt_2024} further highlights that GenAI literacy should adopt an adaptable approach, allowing individuals to understand generative AI based on their needs and enabling them to grow alongside its rapid evolution. 

Current research indicates that although GenAI tools can significantly improve student learning outcomes, the quality of student-generated prompts often falls short, pointing to a pressing need for enhanced GenAI literacy \cite{Lyu_2024}. Additionally, despite the attempts made to integrate GenAI into educational frameworks, there is a notable lack of research examining how students' GenAI literacy impacts their interaction with GenAI-powered tools within LADs \cite{Chiu_2024}. This research gap is particularly significant given the potential of GenAI chatbots to offer rich, contextually relevant explanations that could substantially improve students' comprehension of complex visualisations in LADs. Understanding these interactions could reveal valuable insights into designing more effective GenAI chatbots for LADs. This would ensure that all students, regardless of their initial literacy levels, can fully benefit from these advanced educational technologies, therefore enhancing learners' comprehension of key insights conveyed through LADs. By understanding how varying levels of GenAI literacy impact interaction with GenAI chatbots, educators and researchers can create more equitable learning experiences and ensure that complex learning analytic visualisations are accessible and understandable to all students. This motivates our second and third research questions: \textbf{RQ2}: To what extent is learners' GenAI literacy associated with their \textit{comprehension} of complex analytical visualisations when using a conventional GenAI chatbot compared to a scaffolding GenAI chatbot? \textbf{RQ3}: To what extent do learners’ \textit{interactions} vary based on their level of GenAI literacy when interacting with the conventional and scaffolding chatbots?
\vspace{-5pt}
\section{Methods}

\subsection{Learning Context}
The context and the dataset used to generate visualisations were collected from a healthcare simulation where students, divided into teams of four, played the roles of two primary (blue and red) and two secondary (green and yellow) nurses. The high-fidelity simulation involved a ward set-up with hospital beds, medical equipment, and advanced patient manikins controlled by teachers to simulate varied heart rates and pulses. The simulation aimed to enhance teamwork, communication, and prioritisation skills in clinical emergencies. Multimodal data collected included positioning (x-y coordinates and body orientation via an indoor positioning system), audio (captured by wireless headset microphones), and heart rate (recorded by FitBit Sense wristbands).

\begin{figure*}[h]
    \centering
    \includegraphics[width=1\linewidth]{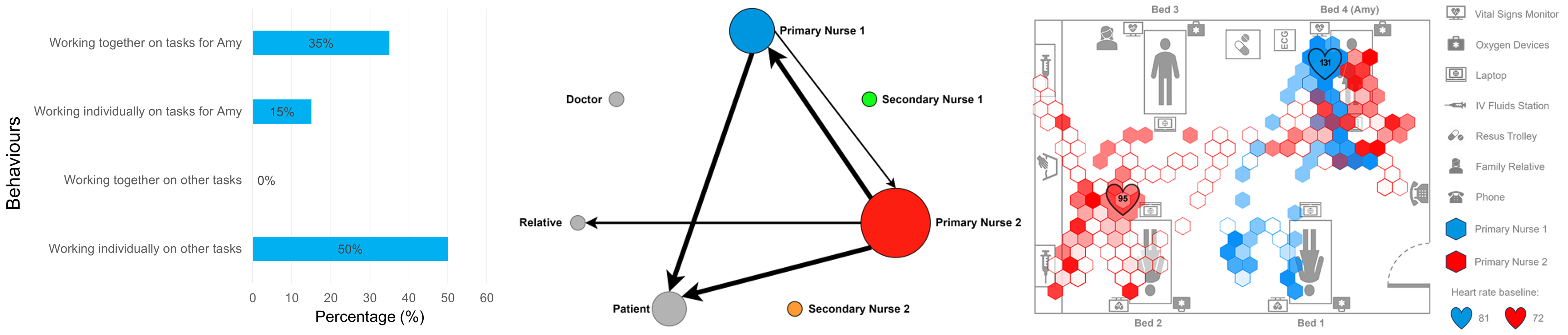}
    \caption{Three LAD visualisations, representing the activity of two nurses in the first phase of their simulation, used in the current study (from left to right): a bar chart, representing the extent of time dedicated to particular tasks; a communication network, representing the amount of conversation among team members; and a ward map showing location, speech and highest heart rate. Enlarged figures are available in our \href{https://osf.io/mwu3x/?view_only=664e0a7fbdf846cc8471df63923681ac}{\textit{repository}}.}
    \label{fig-viz}
\end{figure*}

The multimodal LAD, featuring three visualisations, was developed and validated in previous studies to aid students' post-simulation reflection \cite{martinez2023lessons}. In prior evaluations, students reported difficulties in navigating and understanding the complex visualisations \cite{yan2024evidence}. Consequently, these visualisations offered a chance to explore how GenAI chatbots could enhance comprehension in this study. The three visualisations, as shown in Figure \ref{fig-viz}, vary in complexity: one modality (positioning) for the bar chart, two (positioning and audio) for the communication network, and three (positioning, audio, and heart rate) for the ward map.

The \textbf{bar chart} depicted prioritisation strategies based on students' positional data, allowing easy comparison of time spent on different behaviours and understanding of resource allocation during the simulation \cite{yan2023role}. The \textbf{communication network} or sociogram, mapped interaction patterns using positional and audio data, revealing communication frequencies and directions, including interactions with the patient, doctor, and relative, thus highlighting participant roles and engagement levels \cite{zhao2023mets}. The advanced \textbf{ward map} combined students' physical positions, verbal communication duration, and peak heart rate locations. Inspired by sports analytics \cite{goldsberry2012courtvision}, it used a heatmap to show the frequency and distribution of verbal communications and mapped students' spatial distribution, providing a comprehensive view of physical and behavioural engagement. It also indicated where students experienced the highest physiological arousal by displaying peak heart rate locations and values.

\vspace{-5pt}
\subsection{Design of GenAI chatbots}
\begin{figure*} [h]
    \centering
    \includegraphics[width=1\linewidth]{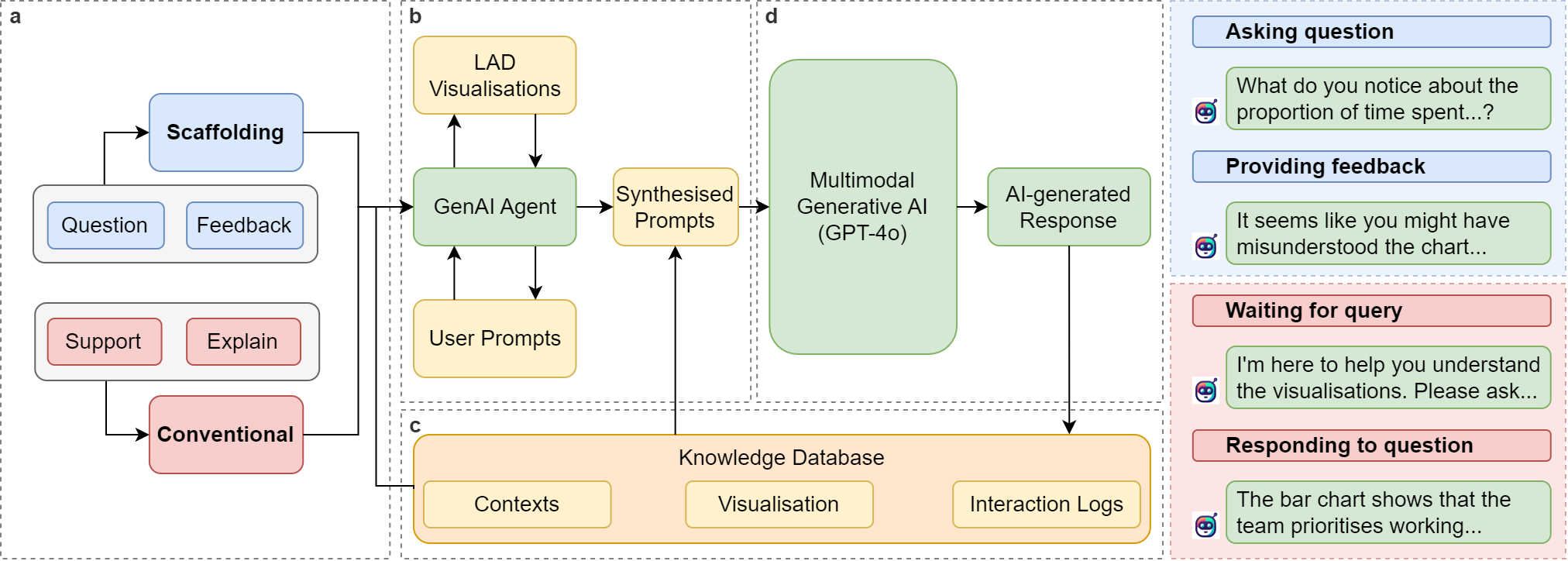}
    \caption{System design of conventional and scaffolding Generative AI (GenAI) chatbots, highlighting four main components: a) distinctive traits separating the conventional chatbot from the scaffolding one, b) interaction dynamics between user prompts, LAD visualisations, and the GenAI agent leveraging retrieval-augmented generation for prompt synthesis, c) a knowledge database with essential contextual information for learning tasks, d) generation of contextually relevant responses using multimodal GenAI. Examples of behaviours: the conventional chatbot (red) focuses on support and explanations, while the scaffolding chatbot (blue) offers guided questions and feedback.}
    \label{fig-genai}
\end{figure*}

To explore how different chatbot designs can support learners with varying levels of GenAI literacy and diverse learning styles, we developed both conventional and scaffolding GenAI chatbots. These chatbots were developed based on the open-source prototype of VizChat \cite{yan2024vizchat}. 
The conventional and scaffolding chatbots were distinguished by their core functionalities and interaction styles (Figure \ref{fig-genai}). \textit{Conventional} chatbots were reactive, delivering precise and contextually relevant information in response to user-initiated queries \cite{ma2023demonstration, yan2024vizchat}, without initiating unsolicited interactions. In contrast, \textit{scaffolding} chatbots were more proactive \cite{park2023generative}, guiding users through visualisations with structured narratives and scaffolding questions crafted by human experts (available in \href{https://osf.io/mwu3x/?view_only=664e0a7fbdf846cc8471df63923681ac}{\textit{repository}}).

\textbf{Prompt integration} is crucial for both types: conventional chatbots processed user queries to generate relevant responses through the interpretation of natural language input and matching it with visualisation and knowledge databases using RAG. Scaffolding chatbots further synthesised prompts based on user interactions and a pre-defined narrative, facilitating exploration with guiding questions and real-time feedback. Both utilised a \textbf{knowledge database} built using LangChain and Chroma, populated with task-specific materials converted into vector embeddings via OpenAI's embedding model (text-embedding-ada-002) \cite{gao2023retrieval}. \textbf{Response generation} leverages GPT-4o's multimodal capabilities: conventional chatbots generated precise answers from semantic searches of the knowledge database, while scaffolding chatbots provided feedback on prior responses, further guiding users through the process step-by-step. For instance, to aid learners in grasping the intricacies of a social network, a scaffolding chatbot could initially pose questions regarding the size of the nodes. Once it had verified that users had a clear understanding through their responses, the chatbot would proceed to query about the edges connecting the nodes. This scaffolding approach, exemplified by guided questions on visualisation elements \cite{xun2004conceptual}, aimed to enhance learners' comprehension of key insights presented by visualisations in LADs. 

\subsection{Study design}

We conducted a 2x2 mixed-method experiment to evaluate the role of GenAI literacy on learner comprehension across baseline and intervention phases. This design also allowed us to analyse learners' cognitive processes when interacting with two types of GenAI chatbots: conventional and scaffolding. The experiment was administered via Qualtrics\footnote{\url{https://www.qualtrics.com/}}, and participants were recruited through Prolific\footnote{\url{https://www.prolific.co/}}. Our sample consisted of current and graduated medical or nursing students with no prior knowledge of the learning context, ensuring improvements were due to the interventions rather than pre-existing familiarity. The study was designed to be completed within an hour, and participants were compensated £8 for their time. Ethics approval was obtained from Monash University (Project Number: 37307), with informed consent from each participant. The study consisted of three main components. The details of each component are elaborated upon as follows:


\subsubsection{Part 1: Demographics and background questions}
First, participants were asked to provide demographic information, including (i) self-reported gender, (ii) age range, (iii) region, and (iv) highest level of education. Following this, they rated their experience in data analysis and GenAI tools on a 5-point Likert scale, ranging from \textit{"None"} (1) to \textit{"Expert"} (5).


\subsubsection{Part 2: Generative AI Literacy Test}
We used a 20-item instrument, the Generative AI Literacy Assessment Test (GLAT) \cite{jin2024glat}, to measure GenAI literacy. Experts reviewed and revised the instrument to ensure face validity, and its content validity was assessed with 204 individuals for relevance, comprehensiveness, and comprehensibility. Structural validity and internal consistency were confirmed using classical test theory and item response theory, resulting in a reliable 2PL model (Cronbach's Alpha = 0.76) and a robust factor structure (\(\chi^2(133) = 135.06\), \(p > .05\), \(\text{RMSEA} = 0.007\), 90\% CI \([0, 0.035]\)). We applied the correction-for-guessing formula \cite{thorndike1991measurement} to account for guessing, as the 2PL model considers only item difficulty and discrimination. The test can be accessed in the \href{https://osf.io/mwu3x/?view_only=664e0a7fbdf846cc8471df63923681ac}{\textit{repository}}.

\subsubsection{Part 3: Comparative Study}
Each participant completed three activities in order: i) reading contextual information about the visualisations, ii) completing a \textit{baseline} analytical writing task and six evaluation questions, and iii) being randomly assigned to one of two \textit{intervention} groups—conventional GenAI chatbots or scaffolding GenAI chatbots—to complete an intervention analytical writing task and six evaluation questions.



The types of visualisations (e.g., bar charts, communication networks, ward maps) remained consistent across both phases, although the specific data and insights varied. The six evaluation questions also maintained a consistent format but were customised to match the insights from each set of visualisations. Additional details on the contextual information, writing tasks, and evaluation questions are provided below.

\textbf{Contextual information.} Participants were provided with a detailed description of a healthcare simulation scenario. In the scenario, two nursing students managed four manikin patients (Beds 1-4) during their shift, focusing primarily on the deteriorating patient in Bed 4 (Amy) while the needs of the other patients also required prioritisation. An actor, playing a relative of the patient in Bed 3, frequently attempted to distract the nurses. The simulation aimed to develop students' skills in teamwork, communication, and prioritisation. This contextual information equipped the participants with necessary background information to understand the visualisations and perform the subsequent writing tasks.


\textbf{Analytical writing tasks.}  
After understanding the contextual information, participants were directed to a custom-built website for the analytical writing tasks, featuring a Learning Analytics Dashboard (LAD). They were asked to analyse three visualisations and wrote a 100-150 word response on \textit{how the two nurses managed the primary patient (Amy) while attending to other beds, focusing on their task prioritisation, verbal communication, and stress levels}. The dashboard contained three components: a \textbf{Display Component}, showing one visualisation at a time to avoid cognitive overload; a \textbf{Writing Space} for articulating analysis; and an \textbf{Instructions Component} with standardised task instructions for the baseline. During the intervention phase, this component was replaced by a \textbf{GenAI Interaction} component. This allowed participants to interact with either conventional or scaffolding GenAI chatbots, depending on their assigned conditions. The task aimed to immerse learners in visual analytics, allow them to comprehend insights, and interact with the chatbot intervention.

\textbf{Evaluation questions.} 
After each analytical writing task, participants answered six multiple-choice questions to assess comprehension of the visualisations. Each visualisation was paired with two questions, designed based on Bloom's taxonomy Levels 1 (knowledge) and 2 (comprehension) \cite{bloom1984bloom}. Knowledge questions assess information retrieval by identifying specific data points or patterns, e.g., \textit{Which behaviour did the two nurses spend the least time on?''} Comprehension questions evaluated interpreting multiple insights and identifying inconsistencies, e.g., \textit{How did the nurses spend their time working on tasks for Amy compared to other tasks?''} Higher taxonomy levels were excluded due to limited contextual knowledge.
Two researchers designed the questions, while a third validated them for applicability across visualisation sets. Discrepancies were resolved through discussion. The question structure remained constant, with randomised answer choices to minimise bias \cite{aera2014}, and included an 'I am not sure' option to reduce guessing \cite{lee2016vlat}. An example of the first set of evaluation questions for the baseline condition is available at the provided \href{https://osf.io/mwu3x/?view_only=664e0a7fbdf846cc8471df63923681ac}{\textit{link}}.


\subsection{Participants}

A total of 81 participants took part in the study, meeting the required sample size (72; 36 per condition) based on a priori power analysis for a medium effect size (0.25) with 80\% power at a 0.05 significance level. Participants were divided into two groups: 41 with a conventional chatbot and 40 with a scaffolding chatbot. All participants were current or graduated students with a medical or nursing background. Participants were from six regions: North/Central America (35), Europe (22), Africa (15), Australia (4), other regions (3), and South America (2). They identified as female (46) and male (35). Ages ranged mostly between 25-34 (32) and 18-24 (29), followed by 35-44 (11), 45-54 (6), 55-64 (2), and 65+ (1). Their education levels were mainly Bachelor's degrees (37) and high school or equivalent (19), with some holding Master's degrees (11), vocational training or diplomas (7), Doctorates (4), or other qualifications (3). Regarding data analysis experience, participants were mostly intermediate (30) or beginners (29), with fewer advanced users (12), no-experience participants (8), and experts (2). Familiarity with GenAI tools was mostly intermediate (44), followed by beginners (25), advanced users (9), experts (2), and one participant with no prior experience.

\subsection{Data Analysis}
First, we defined and calculated three metrics to investigate the association between GenAI literacy and learners' comprehension of complex visualisations using two types of chatbots: a conventional GenAI chatbot and a scaffolding GenAI chatbot. The first metric, \textit{\textbf{Comprehension\_score}}, represents the total number of correct answers out of six evaluation questions, ranging from 0 to 6, with each participant completing two sets of questions to yield \textit{Baseline\_score} and \textit{Intervention\_score}. The \textit{\textbf{Improvement\_score}} quantifies the change in comprehension by calculating the difference between the \textit{Intervention\_score} and the \textit{Baseline\_score}. The \textit{\textbf{GenAI\_literacy}} metric is the sum of correct responses on the GenAI literacy test, ranging from 0 to 20. For both the comprehension score and GenAI literacy score, the correction-for-guessing formula \cite{thorndike1991measurement} was applied to adjust for guessing in these multiple-choice questions with four response options. All calculations and analyses were conducted in Python using libraries such as NumPy, SciPy, and Statsmodels. The next sections provide analysis details for each RQ.

\subsubsection{Preliminary Analysis} 
A preliminary analysis was conducted to ensure the comparability of the two intervention conditions in terms of participants' GenAI literacy, data analysis skills, and GenAI expertise (rated on a scale from 1--None to 5--Expert). This step was crucial to guarantee the robustness and reliability of the subsequent findings \cite{charness2012experimental}. To evaluate whether there were statistically significant differences across the groups, we used the Mann-Whitney test. Additionally, we conducted an epistemic network analysis (ENA) to examine the differences in interactions between participants and GenAI chatbots across the two intervention conditions. This analysis helps determine whether the two groups can be combined into a single dataset for subsequent ENA to address RQ3. Further methodological details of the coding scheme and the ENA procedure are elaborated in section \ref{sec:ENA}.

\subsubsection{RQ1 -- Within-subject Analysis}
To address RQ1, we conducted within-subject analyses to examine changes in learners' comprehension scores for each intervention. We used Wilcoxon signed-rank tests to compare the statistical differences between learners' comprehension scores at baseline and during the intervention, separately for the conventional and scaffolding chatbots. A Mann-Whitney \textit{U} test was conducted to compare the \textit{Improvement\_score} of the two chatbots. Effect sizes were calculated using Rank-Biserial correlation.

\subsubsection{RQ2 -- Regression Analysis (GenAI literacy)}
For RQ2, we used ordinary least squares (OLS) regression to examine the relationship between learners' \textit{GenAI\_literacy} (independent variable; IV) and their \textit{Improvement\_score} (dependent variable; DV). To control for initial comprehension levels, we included the \textit{Baseline\_score} as an additional IV.

\textbf{Formula}: \(\text{\textit{Improvement\_score}} = \beta_0 + \beta_1 \times \text{\textit{Baseline\_score}} + \beta_2 \times \text{\textit{GenAI\_literacy}} + \beta_3 (\text{\textit{Baseline\_score}} \times \text{\textit{GenAI\_literacy}})\)

\noindent This formula included an intercept (\(\beta_0\)), main effects for \textit{Baseline\_score} (\(\beta_1\)) and \textit{GenAI\_literacy} (\(\beta_2\)), and an interaction term (\(\beta_3\)) between \textit{Baseline\_score} and \textit{GenAI\_literacy}. This interaction term helps understand how learners with different levels of GenAI literacy benefit from each intervention. We conducted an ANOVA to decide whether to include the interaction term by comparing models with and without it. The regression analyses were conducted separately for the conventional and scaffolding chatbots to better understand the effects of GenAI literacy on each intervention. The models were fitted using the OLS function from the Statsmodels library, and the assumptions were validated through various methods. To assess linearity, we plotted the predicted values against the observed values. We checked the normality of residuals with the Shapiro-Wilk test and QQ plots. Homoscedasticity was evaluated using the Breusch-Pagan test, and the Durbin-Watson test was used to assess the independence of residuals. All assumptions were met.

\subsubsection{RQ3 -- Epistemic Network Analysis}
\label{sec:ENA}
\textbf{Coding Scheme.}
To analyse user-chatbot interactions within the context of information and cognitive processing, we developed a coding scheme based on Information Processing Theory (IPT) \cite{ATKINSON196889}. IPT aids in understanding how learners process, store, and retrieve information, allowing systematic categorisation and analysis of GenAI chatbot interactions. This approach helps identify how effectively chatbots manage cognitive load, enhance comprehension, and facilitate knowledge transfer. The coding scheme encompassed five constructs: \textit{Information Seeking}, which captures instances where learners or chatbots seek to fill knowledge gaps (input stage); \textit{Clarification}, involving further explanation or rephrasing the information to facilitate comprehension (encoding); \textit{Confirmation}, check the accuracy of their understanding or the information they have stored (verification and validation); \textit{Integration}, integrating new information with existing knowledge (information storage and organisation); and \textit{Reflection}, where individuals reflect on and evaluate processed information (retrieval and metacognitive processes). Each construct is divided into two codes: one for messages from learners (\textit{User}) and one for messages from chatbots (\textit{Chatbot}). Additionally, the code \textit{User.Command} covers other learner commands, such as greetings, signing off, or simple navigation and control commands. Messages were coded at the utterance level, allowing for multiple codes per utterance. Two researchers initially coded 50\% of the data. Cohen’s kappa was used to measure inter-rater reliability, with a threshold of 0.6 for acceptance. Each code achieved a kappa greater than 0.7, indicating strong agreement between coders. One researcher then completed the remaining coding. Users with no chatbot interaction or fewer than two utterances were excluded due to insufficient co-occurrence information. The code \textit{Chatbot.Confirmation} was also removed due to minimal occurrences, resulting in a final set of 10 codes.

\textbf{Epistemic Network Analysis.}
Epistemic Network Analysis (ENA) was conducted to examine the co-occurrence of interaction patterns between learners and GenAI chatbots, segmented by learners' GenAI literacy levels. Learners were categorised into low and high GenAI literacy groups using the Quantile function in Python's Pandas library, with the upper 50\% labelled as high literacy and the remaining 50\% as low literacy. This dichotomous grouping method follows practices in previous ENA studies to model differences in learning behaviours\cite{fan2022dissecting, zhao2023mets}. For the conventional chatbot condition, the analysis included 16 learners with low literacy and 11 with high literacy, while the scaffolding chatbot condition comprised 19 learners with low literacy and 21 with high literacy. In our study, the unit of analysis was defined as the individual interaction episodes between learners and GenAI chatbots, with the utterances from both learners and chatbots treated as lines. We used a stanza window size of two lines to accumulate connections, as recommended for dialogue data analysis \cite{shaffer2016tutorial}. To maximise differences between the mean of the two groups along the x-axis, we used Means Rotation (MR) for dimensional reduction \cite{Bowman2021Math}, facilitating the interpretation of resulting networks. Differences were illustrated by subtracting the mean networks of the two groups in comparison plots. We performed Mann-Whitney \textit{U} tests to examine statistical differences in interaction patterns along the x and y axes, applying Bonferroni correction for multiple comparisons with an initial alpha value of 0.05.

\vspace{-10pt}
\section{Results}

\subsection{Preliminary Findings}

The Mann-Whitney U test for learners' GenAI literacy showed a median score of 11.0 (IQR = 7.0) for the conventional group (n = 41) and 12.0 (IQR = 5.0) for the scaffolding group (n = 40), with no statistically significant difference, \( U = 827, p = .95, r = .01 \). Similarly, for learners' data expertise, both groups had a median score of 3.0 (IQR = 1.0), with no significant difference, \( U = 881, p = .55, r = .06 \). For learners' GenAI expertise, both groups also had median scores of 3.0 (IQR = 1.0), with no significant difference, \( U = 817, p = .98, r = .01 \). These findings indicate that learners in both groups had comparable levels of GenAI literacy, data expertise, and GenAI expertise, supporting both within and between group analyses for RQ1. However, ENA revealed significant differences in learners' interaction patterns between the two groups (see Figure \ref{fig-ena}, left--below in Section \ref{RQ3}). Specifically, a Mann-Whitney test showed that the conventional group (\(Mdn = -0.23, N = 27 \)) was statistically significantly different from the scaffolding group (\(Mdn=0.21, N=40, U=90, p<.001, r =0.83 \)), along the x-axis (MR1; 16.3\%) but not the y-axis (SVD2; 21.1\%). This finding suggests that it is important to analyse these two groups separately to understand specific differences between learners with low and high GenAI literacy (e.g., RQ2 \& 3). 

\subsection{RQ1 -- Comprehension Improvement}

The Mann-Whitney U test for the conventional group indicated significant improvements from the \textit{Baseline\_score} (Median = 3.0, IQR = 1.0) to the \textit{Intervention\_score} (Median = 4.0, IQR = 1.0), \( W = 71, p < .001, r = 0.87 \). This large effect size suggests a substantial increase in comprehension scores due to the conventional GenAI chatbot. Likewise, the Mann-Whitney U test for the scaffolding group also showed significant improvements from the \textit{Baseline\_score} (Median = 3.0, IQR = 2.0) to the \textit{Intervention\_score} (Median = 5.0, IQR = 1.0), \( W = 32, p < .001, r = 0.94 \). This large effect size indicates a substantial improvement in comprehension scores attributed to the scaffolding GenAI chatbot. Additionally, there was no significant difference in the \textit{Improvement\_score} between the two chatbots (\( U = 703, p = .14 \)), indicating that both chatbots improve learners' comprehension. These findings underscore the effectiveness of both the conventional and scaffolding chatbot interventions in enhancing learners' comprehension of complex visualisations.

\subsection{RQ2 -- GenAI Literacy}

For the conventional group, the OLS regression model included the interaction term between \textit{Baseline\_score} and \textit{GenAI\_literacy}, which significantly improved the model's explanatory power (\(\Delta F(1, 38) = 6.27, p = 0.017\)). The model explained 44.5\% of the variance in the \textit{Improvement\_score} (\(F(3, 38) = 10.14, p < .001\)). The intercept was significant (\(\beta = 2.16, SE = 0.48, t = 4.51, p < .001\)), indicating that, on average, learners exhibited an improvement score of 2.16 points when both their \textit{Baseline\_score} and \textit{GenAI\_literacy} were zero. The main effect of \textit{Baseline\_score} was also significant (\(\beta = -0.36, SE = 0.17, t = -2.16, p = 0.037\)), showing that a one-point increase in learners' \textit{Baseline\_score} resulted in a 0.36-point decrease in their \textit{Improvement\_score}. This negative relationship was expected due to the ceiling effect, given the maximum score is six points. The main effect of \textit{GenAI\_literacy} was significant as well (\(\beta = 0.19, SE = 0.07, t = 2.84, p = .007\)), indicating that a one-point increase in learners' \textit{GenAI\_literacy} resulted in a 0.19-point increase in their \textit{Improvement\_score}. This is substantial considering the different scales: the \textit{GenAI\_literacy} score ranged up to 20 points, while the \textit{Improvement\_score} had a mean of 1.1 points and a standard deviation of 1.2 points. Although the interaction term between \textit{Baseline\_score} and \textit{GenAI\_literacy} was significant (\(\beta = -0.05, SE = 0.02, t = -2.51, p = 0.017\)), the effect size was quite small. This indicates a statistically significant but minor negative interaction between \textit{Baseline\_score} and \textit{GenAI\_literacy}. Specifically, for each additional point in \textit{Baseline\_score}, the positive effect of \textit{GenAI\_literacy} on the \textit{Improvement\_score} decreased by 0.05 points, which could also be attributed to the ceiling effect.

For the scaffolding group, the OLS regression model did not include the interaction term, as it did not significantly improve the model's explanatory power (\(\Delta F(1, 37) = 1.02, p = 0.32\)). The model explained 50.2\% of the variance in the \textit{Improvement\_score} (\(F(2, 38) = 19.12, p < .001\)). The intercept was significant (\(\beta = 3.52, SE = 0.46, t = 7.61, p < .001\)), indicating that, on average, learners exhibited an improvement score of 3.52 points when both their \textit{Baseline\_score} and \textit{GenAI\_literacy} are zero. The main effect of \textit{Baseline\_score} was significant (\(\beta = -0.88, SE = 0.14, t = -6.12, p < .001\)), indicating that a one-point increase in learners' \textit{Baseline\_score} resulted in a 0.88-point decrease in their \textit{Improvement\_score}. This negative relationship is consistent with the ceiling effect, given the maximum score is six points. The main effect of \textit{GenAI\_literacy} was also significant, albeit with a relatively small effect size compared to the conventional group (\(\beta = 0.08, SE = 0.03, t = 2.53, p = 0.016\)). This indicates that a one-point increase in learners' \textit{GenAI\_literacy} resulted in a 0.08-point increase in their \textit{Improvement\_score}. 

These findings suggest that higher GenAI literacy positively influences learners' improvement scores in both the conventional and scaffolding chatbot interventions, with a more substantial effect observed in the conventional group.

\subsection{RQ3 -- Cognitive Processing Differences}
\label{RQ3}

\begin{figure*} [h]
    \centering
    \includegraphics[width=1\linewidth]{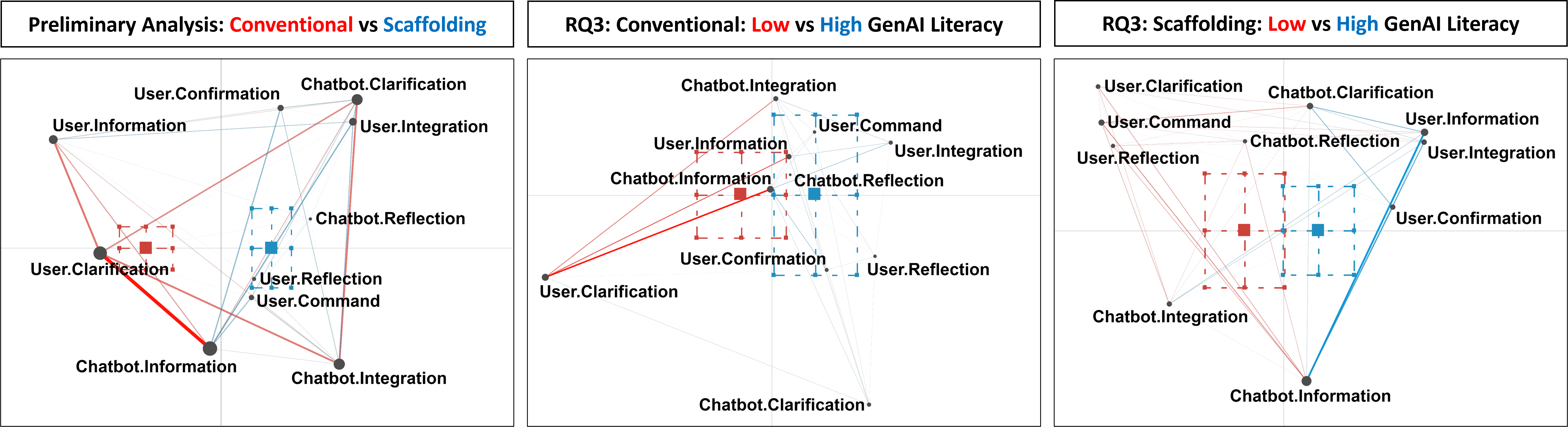}
    \caption{Comparison plots for the preliminary analysis (left), RQ3 conventional (mid), and scaffolding chatbots (right).}
    \label{fig-ena}
\end{figure*}

Within the \textbf{conventional GenAI chatbot} condition, ENA revealed significant differences in interaction patterns between learners with low and high GenAI literacy. A Mann-Whitney test showed that learners with low literacy (Mdn = -0.16, \(N = 16\)) were statistically significantly different from those with high literacy (Mdn = 0.16, \(N = 11, U = 45.00, p = 0.04, r = 0.49\)) along the X-axis (MR1; 16.8\%) with an alpha of 0.05. However, no substantial difference was found along the Y-axis (SVD2; 22.3\%), with the Mann-Whitney test indicating that low GenAI literacy learners (Mdn = 0.08, \(N = 16\)) were not statistically significantly different from high GenAI literacy learners (Mdn = 0.01, \(N = 11; U = 81.00, p = 0.75, r = 0.08\)). As shown in Figure \ref{fig-ena} (mid), learners with low GenAI literacy (red) exhibited a strong emphasis on seeking or offering clarification, as evidenced by the thicker edge between \textit{User.Clarification} and \textit{Chatbot.Information}, reflecting a need for rephrasing and explaining information. Conversely, learners with high GenAI literacy (blue) demonstrated more complex interaction patterns, actively engaging in information exchange fields, such as \textit{User.Integration} and \textit{Chatbot.Information}. They also displayed intricate networks involving \textit{User.Reflection} and \textit{Chatbot.Integration}, indicating a greater propensity to combine new information with prior knowledge and reflect on their understanding.

Within the \textbf{scaffolding GenAI chatbot} condition, ENA also revealed notable differences in interaction patterns between learners with varying levels of GenAI literacy. A Mann-Whitney test indicated that learners with low GenAI literacy (Mdn = -0.09, \(N = 19\)) were statistically significantly different from those with high GenAI literacy (Mdn = 0.12, \(N = 21; U = 308.00, p = 0.00, r = 0.54\)) along the X-axis (MR1; 18.1\%) with an alpha of 0.05. However, similar to the conventional chatbot condition, no significant difference was observed along the Y-axis (SVD2; 26.6\%). The Mann-Whitney test showed that low GenAI literacy learners (Mdn = 0.17, \(N = 19\)) were not statistically significantly different from high GenAI literacy learners (Mdn = 0.13, \(N = 21; U = 217.00, p = 0.64, r = -0.09\)). From Figure \ref{fig-ena} (right), learners with high GenAI literacy (blue) showed the most prominent connection between  \textit{Chabot.Information} and \textit{User.Information}, followed by the connection between Chatbot.Information and \textit{User.Integration}. These connections suggested extensive bidirectional information exchanges and the ability of high-literacy learners to integrate information into their cognitive framework effectively. In contrast, low GenAI literacy learners (red) showed more diverse interactions with the chatbot, exemplified by connections between \textit{User.Clarification}, \textit{User.Command}, \textit{User.Reflection}, and \textit{Chatbot.Information}. These learners often sought clarification, prompted simple commands, and engaged in reflective processes to understand the information before articulating it in writing. As a result, their interactions were less structured and more exploratory, highlighting potential challenges in effectively absorbing and integrating information.


\vspace{-10pt}
\section{Discussion and Conclusion}

This study investigated the effectiveness of conventional and scaffolding chatbots in enhancing learners' comprehension of complex visualisations in LADs and the role of GenAI literacy in these interactions. For \textbf{\textit{RQ1}}, we found that both conventional and scaffolding chatbots significantly improved learners' understanding of insights from complex visualisations in LADs. This finding aligns with and provides empirical support for the proposed benefits of GenAI chatbots \cite{yan2024genai, cukurova2024interplay, khosravi2023generative}. It also underscores the value of transitioning from exploratory to explanatory analytics \cite{echeverria2018driving}, particularly in education, where learners and educators may struggle to extract insights from complex visualisations and therefore, need support and guidance during the process \cite{corrin2018evaluating, Kaliisa2022, fernandez2022beyond}. The current findings advocate for using conventional and scaffolding GenAI chatbots to deliver such support based on their evident effectiveness. Additionally, while both types of chatbots resulted in the same extent of comprehension improvement, the scaffolding chatbots might be a better option for learners with lower GenAI literacy, further discussed in response to RQ2 below.

Regarding \textbf{\textit{RQ2}}, we found that improvements in learners' comprehension are related to their GenAI literacy, with higher literacy leading to greater improvements. This finding aligns with previous research that highlights the importance of considering learners' GenAI literacy in educational studies incorporating GenAI technologies, as the degree of benefit may depend on how effectively learners can utilise the support provided by these technologies \cite{Zhao_2024, Annapureddy_2024, Bozkurt_2024}. Additionally, the stronger positive relationship between GenAI literacy and comprehension improvement with conventional chatbots compared to scaffolding chatbots underscores the necessity of considering GenAI literacy, especially when learners are exposed to GenAI technologies without guidance. This finding resonates with prior studies identifying a low level of prompting skills among students \cite{Lyu_2024, Chiu_2024}. Conversely, the results suggest that integrating GenAI chatbots with scaffolding questions can potentially reduce the dependence on learners' GenAI literacy, making these chatbots broadly applicable regardless of a learner's prior GenAI literacy. This aligns with the documented benefits of scaffolding techniques in supporting better learning \cite{gibbons2002scaffolding, kim2018effectiveness} and contributes to achieving more equitable and inclusive education in the age of GenAI \cite{jin2024generative, yan2024practical}.

In terms of \textbf{\textit{RQ3}}, with the conventional GenAI chatbot, we found that high GenAI literacy learners engaged in more complex cognitive processing activities compared to their lower literacy counterparts. This finding supports the assumption that GenAI literacy might affect a learner's efficiency and effectiveness when engaging with GenAI tools, highlighting the need to measure and foster GenAI literacy as an essential skill \cite{yan2024promises}. For the scaffolding GenAI chatbot condition, high GenAI literacy learners exhibited more structured interactions, whereas low literacy learners displayed more exploratory interactions, indicating they needed additional steps to comprehend and synthesise information into written responses. This aligns with previous studies underscoring the importance of tailored guidance and scaffolding in supporting lower proficiency learners \cite{gibbons2002scaffolding}. Additionally, the finding that low GenAI literacy learners could engage in iterative processes, navigating and learning through exploration and reflection, suggests that the scaffolding chatbot can effectively support these learners by providing dynamic, adaptive assistance to explore and understand complex content \cite{kim2018effectiveness}.

The current study has several \textbf{\textit{implications}}. Specifically, it highlights the need for educational technologies to consider learners' varying levels of GenAI literacy when designing and implementing GenAI-powered tools. Future research should explore strategies to improve GenAI literacy among students \cite{Bozkurt_2024, Lyu_2024, yan2024promises}, thus enabling them to gain greater benefits from these technologies. Additionally, the findings suggest a need for the integration of scaffolding techniques within GenAI chatbots to provide more equitable support across diverse learner populations \cite{gibbons2002scaffolding,  kim2018effectiveness}. Practitioners should consider incorporating scaffolding questions and guided interactions to help learners with lower GenAI literacy effectively navigate complex data visualisations. Furthermore, the study’s findings suggest a promising direction for enhancing LADs by combining explanatory analytics with GenAI-driven personalised support \cite{yan2024vizchat, fernandez2024data}, ultimately contributing to more inclusive and effective educational environments. Continued investigation into the long-term impacts of these technologies on learning outcomes and their potential in various educational contexts will be critical to maximising their utility and ensuring they foster a supportive and equitable learning experience \cite{yan2024promises}.

This study has several \textbf{\textit{limitations}} that should be acknowledged and that pave the way for further research. Firstly, the sample size, while statistically sufficient, was relatively small and may not represent the broader student population, limiting the generalisability of the findings. Future studies should involve larger and more diverse samples to validate these results across different educational contexts. Secondly, the study primarily focused on short-term interactions with GenAI chatbots, and it remains unclear how sustained use over longer periods might affect learners' comprehension and engagement. Longitudinal studies are needed to explore the enduring impacts of GenAI literacy and chatbot integration on learning outcomes. Additionally, while the study underscores the importance of GenAI literacy, it did not address specific instructional strategies for improving this literacy among students. Future research could develop and evaluate targeted interventions aimed at enhancing GenAI literacy. Finally, the complexities of different disciplines and how GenAI chatbots might uniquely support learning in varied subject areas were not examined. Subsequent studies could investigate the applicability and effectiveness of conventional and scaffolding chatbots across different educational domains, offering a nuanced understanding of their role in diverse learning environments.

\vspace{-5pt}
\begin{acks}
This research was in part supported by the Australian Research Council (DP220101209, DP240100069). L.Y.'s work is fully funded by the Digital Health CRC (Cooperative Research Centre). D.G.'s work was, in part, supported by the DHCRC and Defense Advanced Research Projects Agency (DARPA) through the Knowledge Management at Speed and Scale (KMASS) program (HR0011-22-2-0047). The DHCRC is established and supported under the Australian Government's Cooperative Research Centres Program. The U.S. Government is authorised to reproduce and distribute reprints for Governmental purposes notwithstanding any copyright notation thereon. The views and conclusions contained herein are those of the authors and should not be interpreted as necessarily representing the official policies or endorsements, either expressed or implied, of DARPA or the U.S. Government.
\end{acks}

\bibliographystyle{ACM-Reference-Format}
\bibliography{0_reference.bib}


\begin{thebibliography}{74}


\ifx \showCODEN    \undefined \def \showCODEN     #1{\unskip}     \fi
\ifx \showDOI      \undefined \def \showDOI       #1{#1}\fi
\ifx \showISBNx    \undefined \def \showISBNx     #1{\unskip}     \fi
\ifx \showISBNxiii \undefined \def \showISBNxiii  #1{\unskip}     \fi
\ifx \showISSN     \undefined \def \showISSN      #1{\unskip}     \fi
\ifx \showLCCN     \undefined \def \showLCCN      #1{\unskip}     \fi
\ifx \shownote     \undefined \def \shownote      #1{#1}          \fi
\ifx \showarticletitle \undefined \def \showarticletitle #1{#1}   \fi
\ifx \showURL      \undefined \def \showURL       {\relax}        \fi
\providecommand\bibfield[2]{#2}
\providecommand\bibinfo[2]{#2}
\providecommand\natexlab[1]{#1}
\providecommand\showeprint[2][]{arXiv:#2}

\bibitem[Achiam et~al\mbox{.}(2023)]%
        {achiam2023gpt}
\bibfield{author}{\bibinfo{person}{Josh Achiam}, \bibinfo{person}{Steven Adler}, \bibinfo{person}{Sandhini Agarwal}, \bibinfo{person}{Lama Ahmad}, \bibinfo{person}{Ilge Akkaya}, \bibinfo{person}{Florencia~Leoni Aleman}, \bibinfo{person}{Diogo Almeida}, \bibinfo{person}{Janko Altenschmidt}, \bibinfo{person}{Sam Altman}, \bibinfo{person}{Shyamal Anadkat}, {et~al\mbox{.}}} \bibinfo{year}{2023}\natexlab{}.
\newblock \showarticletitle{Gpt-4 technical report}.
\newblock \bibinfo{journal}{\emph{arXiv preprint arXiv:2303.08774}} (\bibinfo{year}{2023}).
\newblock


\bibitem[Al-Doulat et~al\mbox{.}(2020)]%
        {Ahmad2020}
\bibfield{author}{\bibinfo{person}{Ahmad Al-Doulat}, \bibinfo{person}{Nasheen Nur}, \bibinfo{person}{Alireza Karduni}, \bibinfo{person}{Aileen Benedict}, \bibinfo{person}{Erfan Al-Hossami}, \bibinfo{person}{Mary~Lou Maher}, \bibinfo{person}{Wenwen Dou}, \bibinfo{person}{Mohsen Dorodchi}, {and} \bibinfo{person}{Xi Niu}.} \bibinfo{year}{2020}\natexlab{}.
\newblock \showarticletitle{Making Sense of Student Success and Risk Through Unsupervised Machine Learning and Interactive Storytelling}. In \bibinfo{booktitle}{\emph{Artificial Intelligence in Education}}. \bibinfo{publisher}{Springer International Publishing}, \bibinfo{address}{Cham}, \bibinfo{pages}{3--15}.
\newblock
\showISBNx{978-3-030-52237-7}


\bibitem[Alzubi(2024)]%
        {Alzubi_2024}
\bibfield{author}{\bibinfo{person}{Ali Abbas~Falah Alzubi}.} \bibinfo{year}{2024}\natexlab{}.
\newblock \showarticletitle{Generative Artificial Intelligence in the EFL Writing Context: Students’ Literacy in Perspective}.
\newblock \bibinfo{journal}{\emph{Qubahan Academic Journal}} \bibinfo{volume}{4}, \bibinfo{number}{2} (\bibinfo{date}{May} \bibinfo{year}{2024}), \bibinfo{pages}{59–69}.
\newblock
\showISSN{2709-8206}
\urldef\tempurl%
\url{https://doi.org/10.48161/qaj.v4n2a506}
\showDOI{\tempurl}


\bibitem[Annapureddy et~al\mbox{.}(2024)]%
        {Annapureddy_2024}
\bibfield{author}{\bibinfo{person}{Ravinithesh Annapureddy}, \bibinfo{person}{Alessandro Fornaroli}, {and} \bibinfo{person}{Daniel Gatica-Perez}.} \bibinfo{year}{2024}\natexlab{}.
\newblock \showarticletitle{Generative AI Literacy: Twelve Defining Competencies}.
\newblock \bibinfo{journal}{\emph{Digital Government: Research and Practice}} (\bibinfo{date}{Aug.} \bibinfo{year}{2024}).
\newblock
\showISSN{2639-0175}
\urldef\tempurl%
\url{https://doi.org/10.1145/3685680}
\showDOI{\tempurl}


\bibitem[Association et~al\mbox{.}(2014)]%
        {aera2014}
\bibfield{author}{\bibinfo{person}{American Educational~Research Association}, \bibinfo{person}{American~Psychological Association}, {and} \bibinfo{person}{National~Council on~Measurement~in Education}.} \bibinfo{year}{2014}\natexlab{}.
\newblock \bibinfo{booktitle}{\emph{Standards for Educational and Psychological Testing}}.
\newblock \bibinfo{publisher}{AERA}, \bibinfo{address}{Washington, DC}.
\newblock


\bibitem[Atkinson(1968)]%
        {ATKINSON196889}
\bibfield{author}{\bibinfo{person}{Richard~C Atkinson}.} \bibinfo{year}{1968}\natexlab{}.
\newblock \showarticletitle{Human memory: A proposed system and its control processes}.
\newblock \bibinfo{journal}{\emph{The psychology of learning and motivation}}  \bibinfo{volume}{2} (\bibinfo{year}{1968}).
\newblock


\bibitem[Bahroun et~al\mbox{.}(2023)]%
        {bahroun2023transforming}
\bibfield{author}{\bibinfo{person}{Zied Bahroun}, \bibinfo{person}{Chiraz Anane}, \bibinfo{person}{Vian Ahmed}, {and} \bibinfo{person}{Andrew Zacca}.} \bibinfo{year}{2023}\natexlab{}.
\newblock \showarticletitle{Transforming education: A comprehensive review of generative artificial intelligence in educational settings through bibliometric and content analysis}.
\newblock \bibinfo{journal}{\emph{Sustainability}} \bibinfo{volume}{15}, \bibinfo{number}{17} (\bibinfo{year}{2023}), \bibinfo{pages}{12983}.
\newblock


\bibitem[Bloom et~al\mbox{.}(1984)]%
        {bloom1984bloom}
\bibfield{author}{\bibinfo{person}{Benjamin~S Bloom}, \bibinfo{person}{David~R Krathwohl}, \bibinfo{person}{Bertram~B Masia}, {et~al\mbox{.}}} \bibinfo{year}{1984}\natexlab{}.
\newblock \showarticletitle{Bloom taxonomy of educational objectives}.
\newblock In \bibinfo{booktitle}{\emph{Allyn and Bacon}}. \bibinfo{publisher}{Pearson Education London}.
\newblock


\bibitem[Bowman et~al\mbox{.}(2021)]%
        {Bowman2021Math}
\bibfield{author}{\bibinfo{person}{Dale Bowman}, \bibinfo{person}{Zachari Swiecki}, \bibinfo{person}{Zhiqiang Cai}, \bibinfo{person}{Yeyu Wang}, \bibinfo{person}{Brendan Eagan}, \bibinfo{person}{Jeff Linderoth}, {and} \bibinfo{person}{David~Williamson Shaffer}.} \bibinfo{year}{2021}\natexlab{}.
\newblock \showarticletitle{The Mathematical Foundations of Epistemic Network Analysis}. In \bibinfo{booktitle}{\emph{Advances in Quantitative Ethnography}}, \bibfield{editor}{\bibinfo{person}{Andrew~R. Ruis} {and} \bibinfo{person}{Seung~B. Lee}} (Eds.). \bibinfo{publisher}{Springer International Publishing}, \bibinfo{address}{Cham}, \bibinfo{pages}{91--105}.
\newblock
\showISBNx{978-3-030-67788-6}


\bibitem[Bozkurt(2024)]%
        {Bozkurt_2024}
\bibfield{author}{\bibinfo{person}{Aras Bozkurt}.} \bibinfo{year}{2024}\natexlab{}.
\newblock \showarticletitle{Why Generative AI Literacy, Why Now and Why it Matters in the Educational Landscape? Kings, Queens and GenAI Dragons}.
\newblock \bibinfo{journal}{\emph{Open Praxis}} \bibinfo{volume}{16}, \bibinfo{number}{3} (\bibinfo{year}{2024}), \bibinfo{pages}{283–290}.
\newblock
\showISSN{2304-070X}


\bibitem[Charness et~al\mbox{.}(2012)]%
        {charness2012experimental}
\bibfield{author}{\bibinfo{person}{Gary Charness}, \bibinfo{person}{Uri Gneezy}, {and} \bibinfo{person}{Michael~A Kuhn}.} \bibinfo{year}{2012}\natexlab{}.
\newblock \showarticletitle{Experimental methods: Between-subject and within-subject design}.
\newblock \bibinfo{journal}{\emph{Journal of economic behavior \& organization}} \bibinfo{volume}{81}, \bibinfo{number}{1} (\bibinfo{year}{2012}), \bibinfo{pages}{1--8}.
\newblock


\bibitem[Chiu(2024)]%
        {Chiu_2024}
\bibfield{author}{\bibinfo{person}{Thomas~K.F. Chiu}.} \bibinfo{year}{2024}\natexlab{}.
\newblock \showarticletitle{Future research recommendations for transforming higher education with generative AI}.
\newblock \bibinfo{journal}{\emph{Computers and Education: Artificial Intelligence}}  \bibinfo{volume}{6} (\bibinfo{date}{June} \bibinfo{year}{2024}), \bibinfo{pages}{100197}.
\newblock
\showISSN{2666-920X}


\bibitem[Corrin(2018)]%
        {corrin2018evaluating}
\bibfield{author}{\bibinfo{person}{Linda Corrin}.} \bibinfo{year}{2018}\natexlab{}.
\newblock \showarticletitle{Evaluating students’ interpretation of feedback in interactive dashboards}.
\newblock \bibinfo{journal}{\emph{Score reporting research and applications}} (\bibinfo{year}{2018}), \bibinfo{pages}{145--159}.
\newblock


\bibitem[Cukurova(2024)]%
        {cukurova2024interplay}
\bibfield{author}{\bibinfo{person}{Mutlu Cukurova}.} \bibinfo{year}{2024}\natexlab{}.
\newblock \showarticletitle{The interplay of learning, analytics and artificial intelligence in education: A vision for hybrid intelligence}.
\newblock \bibinfo{journal}{\emph{BJET}} (\bibinfo{year}{2024}).
\newblock


\bibitem[Echeverria et~al\mbox{.}(2018a)]%
        {echeverria2018driving}
\bibfield{author}{\bibinfo{person}{Vanessa Echeverria}, \bibinfo{person}{Roberto Martinez-Maldonado}, \bibinfo{person}{Roger Granda}, \bibinfo{person}{Katherine Chiluiza}, \bibinfo{person}{Cristina Conati}, {and} \bibinfo{person}{Simon Buckingham~Shum}.} \bibinfo{year}{2018}\natexlab{a}.
\newblock \showarticletitle{Driving data storytelling from learning design}. In \bibinfo{booktitle}{\emph{Proceedings of the 8th international conference on learning analytics and knowledge}}. \bibinfo{pages}{131--140}.
\newblock


\bibitem[Echeverria et~al\mbox{.}(2018b)]%
        {echeverria2018exploratory}
\bibfield{author}{\bibinfo{person}{Vanessa Echeverria}, \bibinfo{person}{Roberto Martinez-Maldonado}, \bibinfo{person}{Simon~Buckingham Shum}, \bibinfo{person}{Katherine Chiluiza}, \bibinfo{person}{Roger Granda}, {and} \bibinfo{person}{Cristina Conati}.} \bibinfo{year}{2018}\natexlab{b}.
\newblock \showarticletitle{Exploratory versus explanatory visual learning analytics: Driving teachers’ attention through educational data storytelling}.
\newblock \bibinfo{journal}{\emph{JLA}} \bibinfo{volume}{5}, \bibinfo{number}{3} (\bibinfo{year}{2018}), \bibinfo{pages}{73--97}.
\newblock


\bibitem[Fan et~al\mbox{.}(2022)]%
        {fan2022dissecting}
\bibfield{author}{\bibinfo{person}{Yizhou Fan}, \bibinfo{person}{Yuanru Tan}, \bibinfo{person}{Mladen Raković}, \bibinfo{person}{Yeyu Wang}, \bibinfo{person}{Zhiqiang Cai}, \bibinfo{person}{David~Williamson Shaffer}, {and} \bibinfo{person}{Dragan Ga\v{s}evi\'{c}}.} \bibinfo{year}{2022}\natexlab{}.
\newblock \showarticletitle{Dissecting Learning Tactics in {MOOC} using Ordered Network Analysis}.
\newblock \bibinfo{journal}{\emph{Journal of Computer Assisted Learning}} (\bibinfo{year}{2022}), \bibinfo{pages}{in press}.
\newblock


\bibitem[Fernandez~Nieto et~al\mbox{.}(2022)]%
        {fernandez2022beyond}
\bibfield{author}{\bibinfo{person}{Gloria~Milena Fernandez~Nieto}, \bibinfo{person}{Kirsty Kitto}, \bibinfo{person}{Simon Buckingham~Shum}, {and} \bibinfo{person}{Roberto Mart{\'\i}nez-Maldonado}.} \bibinfo{year}{2022}\natexlab{}.
\newblock \showarticletitle{Beyond the learning analytics dashboard: Alternative ways to communicate student data insights combining visualisation, narrative and storytelling}. In \bibinfo{booktitle}{\emph{LAK '22}}. \bibinfo{pages}{219--229}.
\newblock


\bibitem[Fernandez-Nieto et~al\mbox{.}(2024)]%
        {fernandez2024data}
\bibfield{author}{\bibinfo{person}{Gloria~Milena Fernandez-Nieto}, \bibinfo{person}{Roberto Martinez-Maldonado}, \bibinfo{person}{Vanessa Echeverria}, \bibinfo{person}{Kirsty Kitto}, \bibinfo{person}{Dragan Ga{\v{s}}evi{\'c}}, {and} \bibinfo{person}{Simon Buckingham~Shum}.} \bibinfo{year}{2024}\natexlab{}.
\newblock \showarticletitle{Data storytelling editor: A teacher-centred tool for customising learning analytics dashboard narratives}. In \bibinfo{booktitle}{\emph{LAK '24}}. \bibinfo{pages}{678--689}.
\newblock


\bibitem[Gao et~al\mbox{.}(2023)]%
        {gao2023retrieval}
\bibfield{author}{\bibinfo{person}{Yunfan Gao}, \bibinfo{person}{Yun Xiong}, \bibinfo{person}{Xinyu Gao}, \bibinfo{person}{Kangxiang Jia}, \bibinfo{person}{Jinliu Pan}, \bibinfo{person}{Yuxi Bi}, \bibinfo{person}{Yi Dai}, \bibinfo{person}{Jiawei Sun}, {and} \bibinfo{person}{Haofen Wang}.} \bibinfo{year}{2023}\natexlab{}.
\newblock \showarticletitle{Retrieval-augmented generation for large language models: A survey}.
\newblock \bibinfo{journal}{\emph{arXiv preprint arXiv:2312.10997}} (\bibinfo{year}{2023}).
\newblock


\bibitem[Gibbons(2002)]%
        {gibbons2002scaffolding}
\bibfield{author}{\bibinfo{person}{Pauline Gibbons}.} \bibinfo{year}{2002}\natexlab{}.
\newblock \bibinfo{booktitle}{\emph{Scaffolding language, scaffolding learning}}.
\newblock \bibinfo{publisher}{Heinemann Portsmouth, NH}.
\newblock


\bibitem[Goldsberry(2012)]%
        {goldsberry2012courtvision}
\bibfield{author}{\bibinfo{person}{Kirk Goldsberry}.} \bibinfo{year}{2012}\natexlab{}.
\newblock \showarticletitle{Courtvision: New visual and spatial analytics for the nba}. In \bibinfo{booktitle}{\emph{2012 MIT Sloan sports analytics conference}}, Vol.~\bibinfo{volume}{9}. \bibinfo{pages}{12--15}.
\newblock


\bibitem[Gris et~al\mbox{.}(2023)]%
        {gris2023evaluating}
\bibfield{author}{\bibinfo{person}{Lucas Rafael~Stefanel Gris}, \bibinfo{person}{Ricardo Marcacini}, \bibinfo{person}{Arnaldo~Candido Junior}, \bibinfo{person}{Edresson Casanova}, \bibinfo{person}{Anderson Soares}, {and} \bibinfo{person}{Sandra~Maria Alu{\'\i}sio}.} \bibinfo{year}{2023}\natexlab{}.
\newblock \showarticletitle{Evaluating OpenAI's Whisper ASR for Punctuation Prediction and Topic Modeling of life histories of the Museum of the Person}.
\newblock \bibinfo{journal}{\emph{arXiv preprint arXiv:2305.14580}} (\bibinfo{year}{2023}).
\newblock


\bibitem[Hennessy et~al\mbox{.}(2024)]%
        {hennessy2024bjet}
\bibfield{author}{\bibinfo{person}{Sara Hennessy}, \bibinfo{person}{Mutlu Cukurova}, \bibinfo{person}{Cathy Lewin}, \bibinfo{person}{Manolis Mavrikis}, {and} \bibinfo{person}{Louis Major}.} \bibinfo{year}{2024}\natexlab{}.
\newblock \bibinfo{title}{BJET Editorial 2024: A call for research rigour}.
\newblock , \bibinfo{numpages}{5--9}~pages.
\newblock


\bibitem[Ji et~al\mbox{.}(2023)]%
        {ji2023survey}
\bibfield{author}{\bibinfo{person}{Ziwei Ji}, \bibinfo{person}{Nayeon Lee}, \bibinfo{person}{Rita Frieske}, \bibinfo{person}{Tiezheng Yu}, \bibinfo{person}{Dan Su}, \bibinfo{person}{Yan Xu}, \bibinfo{person}{Etsuko Ishii}, \bibinfo{person}{Ye~Jin Bang}, \bibinfo{person}{Andrea Madotto}, {and} \bibinfo{person}{Pascale Fung}.} \bibinfo{year}{2023}\natexlab{}.
\newblock \showarticletitle{Survey of hallucination in natural language generation}.
\newblock \bibinfo{journal}{\emph{Comput. Surveys}} \bibinfo{volume}{55}, \bibinfo{number}{12} (\bibinfo{year}{2023}), \bibinfo{pages}{1--38}.
\newblock


\bibitem[Jin et~al\mbox{.}(2024a)]%
        {jin2024glat}
\bibfield{author}{\bibinfo{person}{Yueqiao Jin}, \bibinfo{person}{Roberto Martinez-Maldonado}, \bibinfo{person}{Dragan Ga{\v{s}}evi{\'c}}, {and} \bibinfo{person}{Lixiang Yan}.} \bibinfo{year}{2024}\natexlab{a}.
\newblock \showarticletitle{GLAT: The Generative AI Literacy Assessment Test}.
\newblock \bibinfo{journal}{\emph{arXiv preprint arXiv:2411.00283}} (\bibinfo{year}{2024}).
\newblock


\bibitem[Jin et~al\mbox{.}(2024b)]%
        {jin2024generative}
\bibfield{author}{\bibinfo{person}{Yueqiao Jin}, \bibinfo{person}{Lixiang Yan}, \bibinfo{person}{Vanessa Echeverria}, \bibinfo{person}{Dragan Ga{\v{s}}evi{\'c}}, {and} \bibinfo{person}{Roberto Martinez-Maldonado}.} \bibinfo{year}{2024}\natexlab{b}.
\newblock \showarticletitle{Generative AI in Higher Education: A Global Perspective of Institutional Adoption Policies and Guidelines}.
\newblock \bibinfo{journal}{\emph{arXiv preprint arXiv:2405.11800}} (\bibinfo{year}{2024}).
\newblock


\bibitem[Kaliisa et~al\mbox{.}(2024)]%
        {Kaliisa_2024}
\bibfield{author}{\bibinfo{person}{Rogers Kaliisa}, \bibinfo{person}{Kamila Misiejuk}, \bibinfo{person}{Sonsoles López-Pernas}, \bibinfo{person}{Mohammad Khalil}, {and} \bibinfo{person}{Mohammed Saqr}.} \bibinfo{year}{2024}\natexlab{}.
\newblock \showarticletitle{Have Learning Analytics Dashboards Lived Up to the Hype? A Systematic Review of Impact on Students’ Achievement, Motivation, Participation and Attitude}. In \bibinfo{booktitle}{\emph{LAK '24}} \emph{(\bibinfo{series}{LAK ’24})}. \bibinfo{publisher}{ACM}.
\newblock


\bibitem[Kaliisa et~al\mbox{.}(2022)]%
        {Kaliisa2022}
\bibfield{author}{\bibinfo{person}{Rogers Kaliisa}, \bibinfo{person}{Anders Mørch}, {and} \bibinfo{person}{Anders Kluge}.} \bibinfo{year}{2022}\natexlab{}.
\newblock \showarticletitle{‘My Point of Departure for Analytics is Extreme Skepticism’: Implications Derived from An Investigation of University Teachers’ Learning Analytics Perspectives and Design Practices}.
\newblock \bibinfo{journal}{\emph{Technology, Knowledge and Learning}}  \bibinfo{volume}{27} (\bibinfo{date}{06} \bibinfo{year}{2022}).
\newblock


\bibitem[Khosravi et~al\mbox{.}(2023)]%
        {khosravi2023generative}
\bibfield{author}{\bibinfo{person}{Hassan Khosravi}, \bibinfo{person}{Olga Viberg}, \bibinfo{person}{Vitomir Kovanovic}, {and} \bibinfo{person}{Rebecca Ferguson}.} \bibinfo{year}{2023}\natexlab{}.
\newblock \showarticletitle{Generative AI and Learning Analytics}.
\newblock \bibinfo{journal}{\emph{JLA}} \bibinfo{volume}{10}, \bibinfo{number}{3} (\bibinfo{year}{2023}), \bibinfo{pages}{1--6}.
\newblock


\bibitem[Kim et~al\mbox{.}(2016)]%
        {kim2016effects}
\bibfield{author}{\bibinfo{person}{Jeonghyun Kim}, \bibinfo{person}{Il-Hyun Jo}, {and} \bibinfo{person}{Yeonjeong Park}.} \bibinfo{year}{2016}\natexlab{}.
\newblock \showarticletitle{Effects of learning analytics dashboard: analyzing the relations among dashboard utilization, satisfaction, and learning achievement}.
\newblock \bibinfo{journal}{\emph{Asia Pacific Education Review}}  \bibinfo{volume}{17} (\bibinfo{year}{2016}), \bibinfo{pages}{13--24}.
\newblock


\bibitem[Kim et~al\mbox{.}(2018)]%
        {kim2018effectiveness}
\bibfield{author}{\bibinfo{person}{Nam~Ju Kim}, \bibinfo{person}{Brian~R Belland}, {and} \bibinfo{person}{Andrew~E Walker}.} \bibinfo{year}{2018}\natexlab{}.
\newblock \showarticletitle{Effectiveness of computer-based scaffolding in the context of problem-based learning for STEM education: Bayesian meta-analysis}.
\newblock \bibinfo{journal}{\emph{Educational Psychology Review}}  \bibinfo{volume}{30} (\bibinfo{year}{2018}), \bibinfo{pages}{397--429}.
\newblock


\bibitem[Kuhail et~al\mbox{.}(2022)]%
        {Kuhail_2022}
\bibfield{author}{\bibinfo{person}{Mohammad~Amin Kuhail}, \bibinfo{person}{Nazik Alturki}, \bibinfo{person}{Salwa Alramlawi}, {and} \bibinfo{person}{Kholood Alhejori}.} \bibinfo{year}{2022}\natexlab{}.
\newblock \showarticletitle{Interacting with educational chatbots: A systematic review}.
\newblock \bibinfo{journal}{\emph{Education and Information Technologies}} \bibinfo{volume}{28}, \bibinfo{number}{1} (\bibinfo{date}{July} \bibinfo{year}{2022}), \bibinfo{pages}{973–1018}.
\newblock
\showISSN{1573-7608}
\urldef\tempurl%
\url{https://doi.org/10.1007/s10639-022-11177-3}
\showDOI{\tempurl}


\bibitem[Lee et~al\mbox{.}(2016)]%
        {lee2016vlat}
\bibfield{author}{\bibinfo{person}{Sukwon Lee}, \bibinfo{person}{Sung-Hee Kim}, {and} \bibinfo{person}{Bum~Chul Kwon}.} \bibinfo{year}{2016}\natexlab{}.
\newblock \showarticletitle{Vlat: Development of a visualization literacy assessment test}.
\newblock \bibinfo{journal}{\emph{IEEE transactions on visualization and computer graphics}} \bibinfo{volume}{23}, \bibinfo{number}{1} (\bibinfo{year}{2016}), \bibinfo{pages}{551--560}.
\newblock


\bibitem[Lee et~al\mbox{.}(2023)]%
        {lee2023prompt}
\bibfield{author}{\bibinfo{person}{Unggi Lee}, \bibinfo{person}{Ariel Han}, \bibinfo{person}{Jeongjin Lee}, \bibinfo{person}{Eunseo Lee}, \bibinfo{person}{Jiwon Kim}, \bibinfo{person}{Hyeoncheol Kim}, {and} \bibinfo{person}{Cheolil Lim}.} \bibinfo{year}{2023}\natexlab{}.
\newblock \showarticletitle{Prompt Aloud!: Incorporating image-generative AI into STEAM class with learning analytics using prompt data}.
\newblock \bibinfo{journal}{\emph{Education and Information Technologies}} (\bibinfo{year}{2023}), \bibinfo{pages}{1--31}.
\newblock


\bibitem[Li et~al\mbox{.}(2024)]%
        {li2024we}
\bibfield{author}{\bibinfo{person}{Haotian Li}, \bibinfo{person}{Yun Wang}, {and} \bibinfo{person}{Huamin Qu}.} \bibinfo{year}{2024}\natexlab{}.
\newblock \showarticletitle{Where are we so far? understanding data storytelling tools from the perspective of human-ai collaboration}. In \bibinfo{booktitle}{\emph{Proceedings of the CHI Conference on Human Factors in Computing Systems}}. \bibinfo{pages}{1--19}.
\newblock


\bibitem[Lim et~al\mbox{.}(2023)]%
        {lim2023effects}
\bibfield{author}{\bibinfo{person}{Lyn Lim}, \bibinfo{person}{Maria Bannert}, \bibinfo{person}{Joep van~der Graaf}, \bibinfo{person}{Shaveen Singh}, \bibinfo{person}{Yizhou Fan}, \bibinfo{person}{Surya Surendrannair}, \bibinfo{person}{Mladen Rakovic}, \bibinfo{person}{Inge Molenaar}, \bibinfo{person}{Johanna Moore}, {and} \bibinfo{person}{Dragan Ga{\v{s}}evi{\'c}}.} \bibinfo{year}{2023}\natexlab{}.
\newblock \showarticletitle{Effects of real-time analytics-based personalized scaffolds on students’ self-regulated learning}.
\newblock \bibinfo{journal}{\emph{Computers in Human Behavior}}  \bibinfo{volume}{139} (\bibinfo{year}{2023}), \bibinfo{pages}{107547}.
\newblock


\bibitem[Long and Magerko(2020)]%
        {Long_2020}
\bibfield{author}{\bibinfo{person}{Duri Long} {and} \bibinfo{person}{Brian Magerko}.} \bibinfo{year}{2020}\natexlab{}.
\newblock \showarticletitle{What is AI Literacy? Competencies and Design Considerations}. In \bibinfo{booktitle}{\emph{CHI ’20}}. \bibinfo{publisher}{ACM}.
\newblock


\bibitem[Lyu et~al\mbox{.}(2024)]%
        {Lyu_2024}
\bibfield{author}{\bibinfo{person}{Wenhan Lyu}, \bibinfo{person}{Yimeng Wang}, \bibinfo{person}{Tingting~(Rachel) Chung}, \bibinfo{person}{Yifan Sun}, {and} \bibinfo{person}{Yixuan Zhang}.} \bibinfo{year}{2024}\natexlab{}.
\newblock \showarticletitle{Evaluating the Effectiveness of LLMs in Introductory Computer Science Education: A Semester-Long Field Study}. In \bibinfo{booktitle}{\emph{L@S ’24}}. \bibinfo{publisher}{ACM}, \bibinfo{pages}{63–74}.
\newblock


\bibitem[Ma et~al\mbox{.}(2023)]%
        {ma2023demonstration}
\bibfield{author}{\bibinfo{person}{Pingchuan Ma}, \bibinfo{person}{Rui Ding}, \bibinfo{person}{Shuai Wang}, \bibinfo{person}{Shi Han}, {and} \bibinfo{person}{Dongmei Zhang}.} \bibinfo{year}{2023}\natexlab{}.
\newblock \showarticletitle{Demonstration of InsightPilot: An LLM-empowered automated data exploration system}.
\newblock \bibinfo{journal}{\emph{arXiv preprint arXiv:2304.00477}} (\bibinfo{year}{2023}).
\newblock


\bibitem[Martinez-Maldonado et~al\mbox{.}(2020)]%
        {martinez2020data}
\bibfield{author}{\bibinfo{person}{Roberto Martinez-Maldonado}, \bibinfo{person}{Vanessa Echeverria}, \bibinfo{person}{Gloria Fernandez~Nieto}, {and} \bibinfo{person}{Simon Buckingham~Shum}.} \bibinfo{year}{2020}\natexlab{}.
\newblock \showarticletitle{From data to insights: A layered storytelling approach for multimodal learning analytics}. In \bibinfo{booktitle}{\emph{Proceedings of the 2020 chi conference on human factors in computing systems}}. \bibinfo{pages}{1--15}.
\newblock


\bibitem[Martinez-Maldonado et~al\mbox{.}(2023)]%
        {martinez2023lessons}
\bibfield{author}{\bibinfo{person}{Roberto Martinez-Maldonado}, \bibinfo{person}{Vanessa Echeverria}, \bibinfo{person}{Gloria Fernandez-Nieto}, \bibinfo{person}{Lixiang Yan}, \bibinfo{person}{Linxuan Zhao}, \bibinfo{person}{Riordan Alfredo}, \bibinfo{person}{Xinyu Li}, \bibinfo{person}{Samantha Dix}, \bibinfo{person}{Hollie Jaggard}, \bibinfo{person}{Rosie Wotherspoon}, {et~al\mbox{.}}} \bibinfo{year}{2023}\natexlab{}.
\newblock \showarticletitle{Lessons learnt from a multimodal learning analytics deployment in-the-wild}.
\newblock \bibinfo{journal}{\emph{ACM Transactions on Computer-Human Interaction}} \bibinfo{volume}{31}, \bibinfo{number}{1} (\bibinfo{year}{2023}), \bibinfo{pages}{1--41}.
\newblock


\bibitem[Mikolov et~al\mbox{.}(2013)]%
        {mikolov2013distributed}
\bibfield{author}{\bibinfo{person}{Tomas Mikolov}, \bibinfo{person}{Ilya Sutskever}, \bibinfo{person}{Kai Chen}, \bibinfo{person}{Greg~S Corrado}, {and} \bibinfo{person}{Jeff Dean}.} \bibinfo{year}{2013}\natexlab{}.
\newblock \showarticletitle{Distributed representations of words and phrases and their compositionality}.
\newblock \bibinfo{journal}{\emph{Advances in neural information processing systems}}  \bibinfo{volume}{26} (\bibinfo{year}{2013}).
\newblock


\bibitem[Ng et~al\mbox{.}(2021)]%
        {Ng_2021}
\bibfield{author}{\bibinfo{person}{Davy Tsz~Kit Ng}, \bibinfo{person}{Jac Ka~Lok Leung}, \bibinfo{person}{Samuel Kai~Wah Chu}, {and} \bibinfo{person}{Maggie~Shen Qiao}.} \bibinfo{year}{2021}\natexlab{}.
\newblock \showarticletitle{Conceptualizing AI literacy: An exploratory review}.
\newblock \bibinfo{journal}{\emph{Computers and Education: Artificial Intelligence}}  \bibinfo{volume}{2} (\bibinfo{year}{2021}), \bibinfo{pages}{100041}.
\newblock
\showISSN{2666-920X}
\urldef\tempurl%
\url{https://doi.org/10.1016/j.caeai.2021.100041}
\showDOI{\tempurl}


\bibitem[Ochoa(2022)]%
        {ochoa_multimodal_2022}
\bibfield{author}{\bibinfo{person}{Xavier Ochoa}.} \bibinfo{year}{2022}\natexlab{}.
\newblock \showarticletitle{Multimodal Learning Analytics - Rationale, Process, Examples, and Direction}.
\newblock In \bibinfo{booktitle}{\emph{The Handbook of Learning Analytics} (\bibinfo{edition}{2} ed.)}, \bibfield{editor}{\bibinfo{person}{Charles Lang}, \bibinfo{person}{George Siemens}, \bibinfo{person}{Alyssa~Friend Wise}, \bibinfo{person}{Dragan Ga\v{s}evi\'{c}}, {and} \bibinfo{person}{Agathe Merceron}} (Eds.). \bibinfo{publisher}{{SoLAR}}, \bibinfo{pages}{54--65}.
\newblock
\showISBNx{978-0-9952408-3-4}


\bibitem[Okonkwo and Ade-Ibijola(2021)]%
        {okonkwo2021chatbots}
\bibfield{author}{\bibinfo{person}{Chinedu~Wilfred Okonkwo} {and} \bibinfo{person}{Abejide Ade-Ibijola}.} \bibinfo{year}{2021}\natexlab{}.
\newblock \showarticletitle{Chatbots applications in education: A systematic review}.
\newblock \bibinfo{journal}{\emph{Comput. Educ.: Artif. Intell.}}  \bibinfo{volume}{2} (\bibinfo{year}{2021}), \bibinfo{pages}{100033}.
\newblock


\bibitem[Ooi et~al\mbox{.}(2023)]%
        {ooi2023potential}
\bibfield{author}{\bibinfo{person}{Keng-Boon Ooi}, \bibinfo{person}{Garry Wei-Han Tan}, \bibinfo{person}{Mostafa Al-Emran}, \bibinfo{person}{Mohammed~A Al-Sharafi}, \bibinfo{person}{Alexandru Capatina}, \bibinfo{person}{Amrita Chakraborty}, \bibinfo{person}{Yogesh~K Dwivedi}, \bibinfo{person}{Tzu-Ling Huang}, \bibinfo{person}{Arpan~Kumar Kar}, \bibinfo{person}{Voon-Hsien Lee}, {et~al\mbox{.}}} \bibinfo{year}{2023}\natexlab{}.
\newblock \showarticletitle{The potential of generative artificial intelligence across disciplines: Perspectives and future directions}.
\newblock \bibinfo{journal}{\emph{Journal of Computer Information Systems}} (\bibinfo{year}{2023}), \bibinfo{pages}{1--32}.
\newblock


\bibitem[Park et~al\mbox{.}(2023)]%
        {park2023generative}
\bibfield{author}{\bibinfo{person}{Joon~Sung Park}, \bibinfo{person}{Joseph O'Brien}, \bibinfo{person}{Carrie~Jun Cai}, \bibinfo{person}{Meredith~Ringel Morris}, \bibinfo{person}{Percy Liang}, {and} \bibinfo{person}{Michael~S Bernstein}.} \bibinfo{year}{2023}\natexlab{}.
\newblock \showarticletitle{Generative agents: Interactive simulacra of human behavior}. In \bibinfo{booktitle}{\emph{Proceedings of the 36th annual acm symposium on user interface software and technology}}. \bibinfo{pages}{1--22}.
\newblock


\bibitem[Pinargote et~al\mbox{.}(2024)]%
        {pinargote2024automating}
\bibfield{author}{\bibinfo{person}{Adriano Pinargote}, \bibinfo{person}{Eddy Calder{\'o}n}, \bibinfo{person}{Kevin Cevallos}, \bibinfo{person}{Gladys Carrillo}, \bibinfo{person}{Katherine Chiluiza}, {and} \bibinfo{person}{Vanessa Echeverr{\'\i}a}.} \bibinfo{year}{2024}\natexlab{}.
\newblock \showarticletitle{Automating Data Narratives in Learning Analytics Dashboards using GenAI.}. In \bibinfo{booktitle}{\emph{LAK Workshops}}. \bibinfo{pages}{150--161}.
\newblock


\bibitem[Pokhrel and Awasthi(2021)]%
        {Pokhrel_2021}
\bibfield{author}{\bibinfo{person}{Jeevan Pokhrel} {and} \bibinfo{person}{Aruna Awasthi}.} \bibinfo{year}{2021}\natexlab{}.
\newblock \bibinfo{booktitle}{\emph{Effectiveness of Dashboard and Intervention Design}}.
\newblock \bibinfo{publisher}{Springer International Publishing}, \bibinfo{pages}{93–116}.
\newblock
\showISBNx{9783030812225}
\showISSN{2662-2130}


\bibitem[Pozdniakov et~al\mbox{.}(2023)]%
        {pozdniakov2023teachers}
\bibfield{author}{\bibinfo{person}{Stanislav Pozdniakov}, \bibinfo{person}{Roberto Martinez-Maldonado}, \bibinfo{person}{Yi-Shan Tsai}, \bibinfo{person}{Vanessa Echeverria}, \bibinfo{person}{Namrata Srivastava}, {and} \bibinfo{person}{Dragan Gasevic}.} \bibinfo{year}{2023}\natexlab{}.
\newblock \showarticletitle{How Do Teachers Use Dashboards Enhanced with Data Storytelling Elements According to their Data Visualisation Literacy Skills?}. In \bibinfo{booktitle}{\emph{LAK'23}}. \bibinfo{pages}{89--99}.
\newblock


\bibitem[Ramaswami et~al\mbox{.}(2022)]%
        {Ramaswami_2022}
\bibfield{author}{\bibinfo{person}{Gomathy Ramaswami}, \bibinfo{person}{Teo Susnjak}, \bibinfo{person}{Anuradha Mathrani}, {and} \bibinfo{person}{Rahila Umer}.} \bibinfo{year}{2022}\natexlab{}.
\newblock \showarticletitle{Use of Predictive Analytics within Learning Analytics Dashboards: A Review of Case Studies}.
\newblock \bibinfo{journal}{\emph{Technology, Knowledge and Learning}} \bibinfo{volume}{28}, \bibinfo{number}{3} (\bibinfo{date}{Aug.} \bibinfo{year}{2022}), \bibinfo{pages}{959–980}.
\newblock
\showISSN{2211-1670}
\urldef\tempurl%
\url{https://doi.org/10.1007/s10758-022-09613-x}
\showDOI{\tempurl}


\bibitem[Ren et~al\mbox{.}(2023)]%
        {ren2023visual}
\bibfield{author}{\bibinfo{person}{Wenqi Ren}, \bibinfo{person}{Yang Tang}, \bibinfo{person}{Qiyu Sun}, \bibinfo{person}{Chaoqiang Zhao}, {and} \bibinfo{person}{Qing-Long Han}.} \bibinfo{year}{2023}\natexlab{}.
\newblock \showarticletitle{Visual semantic segmentation based on few/zero-shot learning: An overview}.
\newblock \bibinfo{journal}{\emph{IEEE/CAA Journal of Automatica Sinica}} (\bibinfo{year}{2023}).
\newblock


\bibitem[Sahin and Ifenthaler(2021)]%
        {sahin2021visualizations}
\bibfield{author}{\bibinfo{person}{Muhittin Sahin} {and} \bibinfo{person}{Dirk Ifenthaler}.} \bibinfo{year}{2021}\natexlab{}.
\newblock \showarticletitle{Visualizations and dashboards for learning analytics: A systematic literature review}.
\newblock \bibinfo{journal}{\emph{Visualizations and dashboards for learning analytics}} (\bibinfo{year}{2021}), \bibinfo{pages}{3--22}.
\newblock


\bibitem[Shaffer et~al\mbox{.}(2016)]%
        {shaffer2016tutorial}
\bibfield{author}{\bibinfo{person}{David~Williamson Shaffer}, \bibinfo{person}{Wesley Collier}, {and} \bibinfo{person}{Andrew~R Ruis}.} \bibinfo{year}{2016}\natexlab{}.
\newblock \showarticletitle{A tutorial on epistemic network analysis: Analyzing the structure of connections in cognitive, social, and interaction data}.
\newblock \bibinfo{journal}{\emph{JLA}} \bibinfo{volume}{3}, \bibinfo{number}{3} (\bibinfo{year}{2016}), \bibinfo{pages}{9--45}.
\newblock


\bibitem[Shao et~al\mbox{.}(2024)]%
        {shao2024data}
\bibfield{author}{\bibinfo{person}{Hongbo Shao}, \bibinfo{person}{Roberto Martinez-Maldonado}, \bibinfo{person}{Vanessa Echeverria}, \bibinfo{person}{Lixiang Yan}, {and} \bibinfo{person}{Dragan Gasevic}.} \bibinfo{year}{2024}\natexlab{}.
\newblock \showarticletitle{Data Storytelling in Data Visualisation: Does it Enhance the Efficiency and Effectiveness of Information Retrieval and Insights Comprehension?}. In \bibinfo{booktitle}{\emph{CHI'24}}. \bibinfo{pages}{1--21}.
\newblock


\bibitem[Shuster et~al\mbox{.}(2021)]%
        {shuster2021retrieval}
\bibfield{author}{\bibinfo{person}{Kurt Shuster}, \bibinfo{person}{Spencer Poff}, \bibinfo{person}{Moya Chen}, \bibinfo{person}{Douwe Kiela}, {and} \bibinfo{person}{Jason Weston}.} \bibinfo{year}{2021}\natexlab{}.
\newblock \showarticletitle{Retrieval augmentation reduces hallucination in conversation}.
\newblock \bibinfo{journal}{\emph{arXiv preprint arXiv:2104.07567}} (\bibinfo{year}{2021}).
\newblock


\bibitem[Siriwardhana et~al\mbox{.}(2023)]%
        {siriwardhana2023improving}
\bibfield{author}{\bibinfo{person}{Shamane Siriwardhana}, \bibinfo{person}{Rivindu Weerasekera}, \bibinfo{person}{Elliott Wen}, \bibinfo{person}{Tharindu Kaluarachchi}, \bibinfo{person}{Rajib Rana}, {and} \bibinfo{person}{Suranga Nanayakkara}.} \bibinfo{year}{2023}\natexlab{}.
\newblock \showarticletitle{Improving the domain adaptation of retrieval augmented generation (RAG) models for open domain question answering}.
\newblock \bibinfo{journal}{\emph{Transactions of the Association for Computational Linguistics}}  \bibinfo{volume}{11} (\bibinfo{year}{2023}), \bibinfo{pages}{1--17}.
\newblock


\bibitem[Susnjak et~al\mbox{.}(2022)]%
        {Susnjak_2022}
\bibfield{author}{\bibinfo{person}{Teo Susnjak}, \bibinfo{person}{Gomathy~Suganya Ramaswami}, {and} \bibinfo{person}{Anuradha Mathrani}.} \bibinfo{year}{2022}\natexlab{}.
\newblock \showarticletitle{Learning analytics dashboard: a tool for providing actionable insights to learners}.
\newblock \bibinfo{journal}{\emph{International Journal of Educational Technology in Higher Education}} \bibinfo{volume}{19}, \bibinfo{number}{1} (\bibinfo{date}{Feb.} \bibinfo{year}{2022}).
\newblock
\showISSN{2365-9440}
\urldef\tempurl%
\url{https://doi.org/10.1186/s41239-021-00313-7}
\showDOI{\tempurl}


\bibitem[Thorndike et~al\mbox{.}(1991)]%
        {thorndike1991measurement}
\bibfield{author}{\bibinfo{person}{Robert~M Thorndike}, \bibinfo{person}{George~K Cunningham}, \bibinfo{person}{Robert~Ladd Thorndike}, {and} \bibinfo{person}{Elizabeth~P Hagen}.} \bibinfo{year}{1991}\natexlab{}.
\newblock \bibinfo{booktitle}{\emph{Measurement and evaluation in psychology and education}}.
\newblock \bibinfo{publisher}{Macmillan Publishing Co, Inc}.
\newblock


\bibitem[Verbert et~al\mbox{.}(2020)]%
        {verbert2020learning}
\bibfield{author}{\bibinfo{person}{Katrien Verbert}, \bibinfo{person}{Xavier Ochoa}, \bibinfo{person}{Robin De~Croon}, \bibinfo{person}{Raphael~A Dourado}, {and} \bibinfo{person}{Tinne De~Laet}.} \bibinfo{year}{2020}\natexlab{}.
\newblock \showarticletitle{Learning analytics dashboards: the past, the present and the future}. In \bibinfo{booktitle}{\emph{Proceedings of the 10th Learning Analytics and Knowledge Conference}}. \bibinfo{pages}{35--40}.
\newblock


\bibitem[Wen et~al\mbox{.}(2024)]%
        {wen2024learning}
\bibfield{author}{\bibinfo{person}{Cai-Ting Wen}, \bibinfo{person}{Chen-Chung Liu}, \bibinfo{person}{Ching-Yuan Li}, \bibinfo{person}{Ming-Hua Chang}, \bibinfo{person}{Shih-Hsun~Fan Chiang}, \bibinfo{person}{Hung-Ming Lin}, \bibinfo{person}{Fu-Kwun Hwang}, {and} \bibinfo{person}{Gautam Biswas}.} \bibinfo{year}{2024}\natexlab{}.
\newblock \showarticletitle{The learning analytics of computational scientific modeling with self-explanation for subgoals and demonstration scaffolding}.
\newblock \bibinfo{journal}{\emph{Computers \& Education}}  \bibinfo{volume}{215} (\bibinfo{year}{2024}), \bibinfo{pages}{105043}.
\newblock


\bibitem[White et~al\mbox{.}(2023)]%
        {white2023prompt}
\bibfield{author}{\bibinfo{person}{Jules White}, \bibinfo{person}{Quchen Fu}, \bibinfo{person}{Sam Hays}, \bibinfo{person}{Michael Sandborn}, \bibinfo{person}{Carlos Olea}, \bibinfo{person}{Henry Gilbert}, \bibinfo{person}{Ashraf Elnashar}, \bibinfo{person}{Jesse Spencer-Smith}, {and} \bibinfo{person}{Douglas~C Schmidt}.} \bibinfo{year}{2023}\natexlab{}.
\newblock \showarticletitle{A prompt pattern catalog to enhance prompt engineering with chatgpt}.
\newblock \bibinfo{journal}{\emph{arXiv preprint arXiv:2302.11382}} (\bibinfo{year}{2023}).
\newblock


\bibitem[Wu et~al\mbox{.}(2023)]%
        {wu2023autogen}
\bibfield{author}{\bibinfo{person}{Qingyun Wu}, \bibinfo{person}{Gagan Bansal}, \bibinfo{person}{Jieyu Zhang}, \bibinfo{person}{Yiran Wu}, \bibinfo{person}{Shaokun Zhang}, \bibinfo{person}{Erkang Zhu}, \bibinfo{person}{Beibin Li}, \bibinfo{person}{Li Jiang}, \bibinfo{person}{Xiaoyun Zhang}, {and} \bibinfo{person}{Chi Wang}.} \bibinfo{year}{2023}\natexlab{}.
\newblock \showarticletitle{Autogen: Enabling next-gen llm applications via multi-agent conversation framework}.
\newblock \bibinfo{journal}{\emph{arXiv preprint arXiv:2308.08155}} (\bibinfo{year}{2023}).
\newblock


\bibitem[Xun and Land(2004)]%
        {xun2004conceptual}
\bibfield{author}{\bibinfo{person}{GE Xun} {and} \bibinfo{person}{Susan~M Land}.} \bibinfo{year}{2004}\natexlab{}.
\newblock \showarticletitle{A conceptual framework for scaffolding III-structured problem-solving processes using question prompts and peer interactions}.
\newblock \bibinfo{journal}{\emph{Educational technology research and development}} \bibinfo{volume}{52}, \bibinfo{number}{2} (\bibinfo{year}{2004}), \bibinfo{pages}{5--22}.
\newblock


\bibitem[Yan et~al\mbox{.}(2024a)]%
        {yan2024evidence}
\bibfield{author}{\bibinfo{person}{Lixiang Yan}, \bibinfo{person}{Vanessa Echeverria}, \bibinfo{person}{Yueqiao Jin}, \bibinfo{person}{Gloria Fernandez-Nieto}, \bibinfo{person}{Linxuan Zhao}, \bibinfo{person}{Xinyu Li}, \bibinfo{person}{Riordan Alfredo}, \bibinfo{person}{Zachari Swiecki}, \bibinfo{person}{Dragan Ga{\v{s}}evi{\'c}}, {and} \bibinfo{person}{Roberto Martinez-Maldonado}.} \bibinfo{year}{2024}\natexlab{a}.
\newblock \showarticletitle{Evidence-based multimodal learning analytics for feedback and reflection in collaborative learning}.
\newblock \bibinfo{journal}{\emph{British Journal of Educational Technology}} \bibinfo{volume}{55}, \bibinfo{number}{5} (\bibinfo{year}{2024}), \bibinfo{pages}{1900--1925}.
\newblock


\bibitem[Yan et~al\mbox{.}(2024b)]%
        {yan2024promises}
\bibfield{author}{\bibinfo{person}{Lixiang Yan}, \bibinfo{person}{Samuel Greiff}, \bibinfo{person}{Ziwen Teuber}, {and} \bibinfo{person}{Dragan Ga{\v{s}}evi{\'c}}.} \bibinfo{year}{2024}\natexlab{b}.
\newblock \showarticletitle{Promises and challenges of generative artificial intelligence for human learning}.
\newblock \bibinfo{journal}{\emph{Nature Human Behaviour}} \bibinfo{volume}{8}, \bibinfo{number}{10} (\bibinfo{year}{2024}), \bibinfo{pages}{1839--1850}.
\newblock


\bibitem[Yan et~al\mbox{.}(2024c)]%
        {yan2024genai}
\bibfield{author}{\bibinfo{person}{Lixiang Yan}, \bibinfo{person}{Roberto Martinez-Maldonado}, {and} \bibinfo{person}{Dragan Gasevic}.} \bibinfo{year}{2024}\natexlab{c}.
\newblock \showarticletitle{Generative Artificial Intelligence in Learning Analytics: Contextualising Opportunities and Challenges through the Learning Analytics Cycle}. In \bibinfo{booktitle}{\emph{LAK '24}}. \bibinfo{pages}{101–111}.
\newblock


\bibitem[Yan et~al\mbox{.}(2023)]%
        {yan2023role}
\bibfield{author}{\bibinfo{person}{Lixiang Yan}, \bibinfo{person}{Roberto Martinez-Maldonado}, \bibinfo{person}{Linxuan Zhao}, \bibinfo{person}{Samantha Dix}, \bibinfo{person}{Hollie Jaggard}, \bibinfo{person}{Rosie Wotherspoon}, \bibinfo{person}{Xinyu Li}, {and} \bibinfo{person}{Dragan Ga{\v{s}}evi{\'c}}.} \bibinfo{year}{2023}\natexlab{}.
\newblock \showarticletitle{The role of indoor positioning analytics in assessment of simulation-based learning}.
\newblock \bibinfo{journal}{\emph{BJET}} \bibinfo{volume}{54}, \bibinfo{number}{1} (\bibinfo{year}{2023}), \bibinfo{pages}{267--292}.
\newblock


\bibitem[Yan et~al\mbox{.}(2024d)]%
        {yan2024practical}
\bibfield{author}{\bibinfo{person}{Lixiang Yan}, \bibinfo{person}{Lele Sha}, \bibinfo{person}{Linxuan Zhao}, \bibinfo{person}{Yuheng Li}, \bibinfo{person}{Roberto Martinez-Maldonado}, \bibinfo{person}{Guanliang Chen}, \bibinfo{person}{Xinyu Li}, \bibinfo{person}{Yueqiao Jin}, {and} \bibinfo{person}{Dragan Ga{\v{s}}evi{\'c}}.} \bibinfo{year}{2024}\natexlab{d}.
\newblock \showarticletitle{Practical and ethical challenges of large language models in education: A systematic scoping review}.
\newblock \bibinfo{journal}{\emph{British Journal of Educational Technology}} \bibinfo{volume}{55}, \bibinfo{number}{1} (\bibinfo{year}{2024}), \bibinfo{pages}{90--112}.
\newblock


\bibitem[Yan et~al\mbox{.}(2024e)]%
        {yan2024vizchat}
\bibfield{author}{\bibinfo{person}{Lixiang Yan}, \bibinfo{person}{Linxuan Zhao}, \bibinfo{person}{Vanessa Echeverria}, \bibinfo{person}{Yueqiao Jin}, \bibinfo{person}{Riordan Alfredo}, \bibinfo{person}{Xinyu Li}, \bibinfo{person}{Dragan Ga{\v{s}}evi’c}, {and} \bibinfo{person}{Roberto Martinez-Maldonado}.} \bibinfo{year}{2024}\natexlab{e}.
\newblock \showarticletitle{VizChat: Enhancing Learning Analytics Dashboards with Contextualised Explanations Using Multimodal Generative AI Chatbots}. In \bibinfo{booktitle}{\emph{International Conference on Artificial Intelligence in Education}}. Springer, \bibinfo{pages}{180--193}.
\newblock


\bibitem[Yousef and Khatiry(2021)]%
        {Yousef_2021}
\bibfield{author}{\bibinfo{person}{Ahmed Mohamed~Fahmy Yousef} {and} \bibinfo{person}{Ahmed~Ramadan Khatiry}.} \bibinfo{year}{2021}\natexlab{}.
\newblock \showarticletitle{Cognitive versus behavioral learning analytics dashboards for supporting learner’s awareness, reflection, and learning process}.
\newblock \bibinfo{journal}{\emph{Interactive Learning Environments}} \bibinfo{volume}{31}, \bibinfo{number}{9} (\bibinfo{date}{Dec.} \bibinfo{year}{2021}), \bibinfo{pages}{5460–5476}.
\newblock
\showISSN{1744-5191}


\bibitem[Zhao et~al\mbox{.}(2023)]%
        {zhao2023mets}
\bibfield{author}{\bibinfo{person}{Linxuan Zhao}, \bibinfo{person}{Zachari Swiecki}, \bibinfo{person}{Dragan Ga\v{s}evi\'{c}}, \bibinfo{person}{Lixiang Yan}, \bibinfo{person}{Samantha Dix}, \bibinfo{person}{Hollie Jaggard}, \bibinfo{person}{Rosie Wotherspoon}, \bibinfo{person}{Abra Osborne}, \bibinfo{person}{Xinyu Li}, \bibinfo{person}{Riordan Alfredo}, {and} \bibinfo{person}{Roberto Martinez-Maldonado}.} \bibinfo{year}{2023}\natexlab{}.
\newblock \showarticletitle{{METS}: Multimodal Learning Analytics of Embodied Teamwork Learning}. In \bibinfo{booktitle}{\emph{{LAK '23}}}. \bibinfo{pages}{186–196}.
\newblock


\bibitem[Zhao et~al\mbox{.}(2024)]%
        {Zhao_2024}
\bibfield{author}{\bibinfo{person}{Xin Zhao}, \bibinfo{person}{Andrew Cox}, {and} \bibinfo{person}{Liang Cai}.} \bibinfo{year}{2024}\natexlab{}.
\newblock \showarticletitle{ChatGPT and the digitisation of writing}.
\newblock \bibinfo{journal}{\emph{Humanit. Soc. Sci. Commun.}} \bibinfo{volume}{11}, \bibinfo{number}{1} (\bibinfo{date}{April} \bibinfo{year}{2024}).
\newblock
\showISSN{2662-9992}


\end{thebibliography}






\end{document}